\documentclass[structabstract, print]{aa}
\usepackage{txfonts}
\usepackage{graphicx}

\usepackage{amssymb}
\usepackage{graphicx}
\usepackage{dcolumn}
\usepackage{bm}
\usepackage{amsfonts}
\usepackage{latexsym}
\usepackage{pdfsync}
\usepackage{color}

\begin{document}

\newcommand{\commentout}[1]{}
\newcommand{\nwc}{\newcommand}
\nwc{\mc}{\mathcal}
\nwc{\mb}{\mathbf}
\nwc{\mbb}{\mathbb}
\nwc{\bld}{\textbf}
\nwc{\ti}{\textit}
\nwc{\tu}{\underline}
\nwc{\beq}{\begin{eqnarray}}
\nwc{\eeq}{\end{eqnarray}}
\nwc{\beqn}{\begin{eqnarray*}}
\nwc{\eeqn}{\end{eqnarray*}}
\nwc{\tbf}{\textbf}
\nwc{\cd}{\cdot}
\nwc{\disp}{\displaystyle}

\title{Optimal Arrays for Compressed Sensing in Snapshot-Mode Radio Interferometry}
\author{Clara Fannjiang}
\institute{Stanford University, 353 Serra Mall, Stanford CA 94305, United States}
\date{Received date / Accepted date}

\abstract {Radio interferometry has always faced the problem of incomplete sampling of the Fourier plane. A possible remedy can be found in the promising new theory of compressed sensing (CS), which allows for the accurate recovery of sparse signals from sub-Nyquist sampling given certain measurement conditions.} {We provide an introductory assessment of optimal arrays for CS in snapshot-mode radio interferometry, using orthogonal matching pursuit (OMP), a widely used CS recovery algorithm similar in some respects to CLEAN. We focus on centrally condensed (specifically, Gaussian) arrays versus uniform arrays, and the principle of randomization versus deterministic arrays such as the VLA.} {The theory of CS is grounded in $a)$ sparse representation of signals and $b)$ measurement matrices of low coherence. We calculate a related quantity, mutual coherence (MC), as a theoretical indicator of arrays' suitability for OMP based on the recovery error bounds in (Donoho et al. 2006). OMP reconstructions of both point and extended objects are also run from simulated incomplete data. Optimal arrays are considered for object recovery through 1) the natural pixel representation and 2) the representation by the block discrete cosine transform (BDCT).} {We find that reconstructions of the pixel representation perform best with the uniform random array, while reconstructions of the BDCT representation perform best with normal random arrays. Slight randomization to the VLA also improves it hugely for CS with the pixel basis. } {In the pixel basis, array design for CS reflects known principles of array design for small numbers of antennas, namely of randomness and uniform distribution. Differing results with the BDCT, however, emphasize the importance of studying how sparsifying bases affect array design before CS can be optimized.}


\titlerunning{Arrays for Radio Interferometry by Compressed Sensing}
\authorrunning{C. Fannjiang}
\maketitle

\section{Introduction}
Interferometry is the definitive imaging tool of radio astronomy, probing high-resolution structures by using an array of many antennas to emulate a single lens whose aperture is the greatest distance between a pair of antennas (Thompson et. al. 2001). By measuring the visibility, or the interference fringes of the radio signal at every pair of antennas, an array samples the two-dimensional Fourier transform of the spatial intensity distribution of the source. Ideally, if we thoroughly sample the Fourier, or \ti{u-v}, plane, we can then invert the transform to reconstruct the object. The bulk of data required has motivated the new generation of ambitious interferometers, including the Atacama Large Millimetre/submillimetre Array (ALMA) and the Square Kilometre Array (SKA), which will use several thousand antennas. Meanwhile, smaller interferometers often sample over a period of time, allowing the rotation of the earth to naturally produce new baselines. 

Despite such measures, there are always irregular holes on the \ti{u-v} plane where sampling of the visibility function is thin or simply nonexistent. This data deficiency is currently managed by interpolating or filling in zeros for unknown visibility values, and applying deconvolution algorithms such as CLEAN and its variants (H\"ogbom 1974; Clark 1980) to the resulting Òdirty imagesÓ. However, it may not be necessary to collect the extensive data set in the first place. 

Among signal processing's most promising developments in recent years is the theory of compressed sensing (CS), which has shown that the information of a signal can be preserved even when sampling does not fulfill the fundamental Nyquist rate (Donoho 2006; Cand\`es et al. 2006a, b). The theory revolves around \ti{a priori} knowledge that the signal is sparse or compressible in some basis, in which case its information naturally resides in a relatively small number of coefficients. Instead of directly sampling the signal, whereby full sampling would be inevitable in finding every non-zero or significant coefficient, CS allows us to compute just a few inner products of the signal along selected measurement vectors of certain favorable characteristics. The novelty of CS is that it takes advantage of signal compressibility to alleviate expensive data acquisition, not just data storage. 

It is well established that objects typical of astronomic study are sparse or compressible (Polygiannakis 2003; Dollet 2004); indeed, they are often sparse in the natural pixel basis. Recent studies have recognized this innate agreeability with CS, examining its potential in radio interferometry compared to traditional deconvolution methods (Wiaux et al. 2009; Li et. al. 2011) as well as methods for applying it to wide field-of-view interferometry (McEwen \& Wiaux 2011). As in traditional interferometry, however, reliable imaging will depend heavily on the sample distribution the array produces on the \ti{u-v} plane. CS involves unusual premises for sampling, e.g., that we actively purpose to undersample, and rethinking array design to complement these premises is a critical step before CS can be applied in practice.

In this paper, we search for idealized arrays that optimize the performance of a CS recovery algorithm known as orthogonal matching pursuit (OMP) (Davis et al. 1997; Pati et al. 1993; Tropp \& Gilbert 2007) for snapshot-mode observations when objects are sparsely represented $1)$ in the pixel basis and $2)$ by the block discrete cosine transform (BDCT). We focus especially on how array configuration affects measurement matrix incoherence, a cornerstone of CS performance guarantees, and also simulate reconstructions of both point and extended objects. Among the tested arrays, we find that a uniform random distribution of antennas performs best in the pixel basis and that a normal random distribution performs best with the BDCT. (Initial results with the pixel basis were also briefly summarized in (Fannjiang 2011).) We also study the benefits of slight perturbation to the VLA's patterned 'Y' array, as well as the unexpected inability of measurement matrix incoherence to predict OMP reconstruction behavior with the BDCT. 

In Sect. \ref{CS}, we give a brief description of the CS framework as it pertains to radio interferometry and array optimization. The arrays under study are summarized in Sect. \ref{Arrays}. Using the pixel basis, Sect. \ref{WellRes} analyzes how these arrays affect the incoherence of the measurement matrix, as well as OMP reconstructions of point and extended objects; Sect. \ref{SuperRes} does the same in under-resolved conditions. Section \ref{BDCT} adopts the BDCT instead of the pixel basis to sparsely represent objects, also examining measurement matrix incoherence and extended object reconstructions. The conclusions are drawn in Sect. \ref{Conc}.

\section{Compressed sensing}
\label{CS}
\subsection{Overview}
Consider a signal $\mb x \in {\mbb R}^N$, which has a sparse representation $\mb s$ with the columns of an $N \times N$ basis matrix $\Psi$, such that ${\mb x} = \Psi \mb s$. Given that $\mb s$ is $S$-sparse, meaning it has only $S \ll N$ non-zero components (or, more generally, given that $\mb s$ is $S$-compressible and has only $S \ll N$ significant components) and given measurement vectors of certain desirable characteristics, CS proposes that we should not have to take the complete set of $N$ inner-product measurements to recover the signal.  If  $\Theta$ is an $M \times N$ matrix whose $M < N$ rows are the measurement vectors, then CS aims to invert the underdetermined linear system ${\mb y} = \Theta \Psi {\mb s} = \Phi {\mb s}$, where $\mb y$ is a vector of the $M < N$ measurements of $\mb x$ and we call $\Phi$ the measurement matrix. 

If $\mb s$ is sparse, it follows that we seek a sparse solution to the inverse problem. Indeed, in the absence of noise $\mb s$ is the sparsest solution, i.e., $\mb{\hat s}$ subject to ${\mb y} = \Phi \mb{\hat s}$ with the least number of non-zero components. A direct search for this solution, however, is computationally intractable. Given $\Phi$ of favorable characteristics (described below), an $\ell_1$-minimization problem can be solved instead, such that we search for the solution $\mb{\hat s}$ with the smallest $\ell_1$-norm, or sum of the absolute values of the coefficients. CS recovery schemes also include greedy $\ell_0$-norm minimization methods, such as OMP, which we use in this paper. Faster than basis pursuit (BP) (Chen et al. 1992) and LASSO (Tibshirani 1996), the classic $\ell_1$-minimization methods, OMP was also chosen for its parallels to CLEAN: both iteratively select point sources in a greedy fashion until the residual falls below some stopping threshold. The one difference is that OMP calculates each new residual by subtracting an orthogonal projection of the data onto all columns of $\Phi$ selected so far, rather than simply subtracting the visibility of the point source selected in that iteration (see Sect. \ref{OMP}, and (Pati et al. 1993) and (Davis et al. 1997) for a rigorous presentation).

Recovering $\mb x$ requires that a measurement matrix $\Phi = \Theta \Psi$ is selected that does not corrupt or lose key features of $\mb s$ in mapping higher-dimension $\mb s$ to the lower-dimension $\mb y$. An intuitive ideal is that $\Phi$ should nearly preserve the Euclidean norm of $\mb s$, which means that every possible subset of $S$ columns of $\Phi$ acts as an orthonormal system. A rigorous formulation of such a quality is known as the restricted isometry property (RIP) (Cand\`es \& Tao 2005), but it cannot be verified empirically; checking every $S$-combination of columns of $\Phi$ for near-orthogonality is a combinatorial process and $NP$-hard. An alternative metric that we use is mutual coherence (MC), which can guarantee the RIP (a relationship derived in (Davenport et al. 2012) from results in (Ger\u{s}gorin 1931)) and is well-suited for numerical experimentation. MC gives the maximum correlation between any pair of columns of $\Phi$:

\beq
\label{MC}
\mu (\Phi) = \max_{1 \leq k \neq k \prime \leq N} \frac{|\Phi_k \cdot \Phi_{k \prime}|}{\| \Phi_k\|_2 \| \Phi_{k \prime}\|_2}
\eeq

An incoherent matrix with a near-zero MC is desired. Error bounds for the performance of the major CS recovery processes, including BP (Donoho et al. 2006), LASSO (Cand\`es \& Plan 2009), and OMP (Donoho et al. 2006) have been developed based on MC. For OMP, suppose the sparsity $S$ of $\mb s$ satisfies

\beq
\label{DonohoCond}
S \leq \frac{1 + \mu - 2n}{2 \mu},
\eeq

where $n$ is the ratio of $\epsilon$, the norm of the noise in the measurements, to the absolute value of the least non-zero component of $\mb s$. Then OMP is guaranteed to find the solution $\mb{\hat{s}}$ such that 

\beq
\label{Donoho1}
supp(\mb{\hat{s}}) = supp(\mb{s}),
\eeq

where $supp$ gives the support of its argument, i.e., the indices of the non-zero components of $\mb{\hat{s}}$, and 

\beq
\label{Donoho2}
\|  \mb{\hat{s}} - \mb{s} \|^{2}_{2} \leq  \frac{{\epsilon}^2}{1-\mu(S - 1)}. 
\eeq

as proven by (Donoho et al. 2006). These bounds use $\epsilon$ as the stopping criterion for OMP, i.e., the recovery process stops when $||\mb{V} - \Phi \mb{\hat{I}}||_2 \leq \epsilon$ where $\mb{\hat{I}}$ is the reconstruction of $\mb{I}$. An important note is that these error bounds were developed to encompass all possible $\Phi$ and $\mb s$ combinations, regardless of how ill-suited they are for CS. In practice, when the measurement scheme and object sparsity are tailored for CS application, OMP outperforms these bounds dramatically. In this paper, we thus refer to these bounds as a general representation of the dependence of CS algorithms on measurement matrix incoherence, rather than as a strict prediction of error in reconstruction.

\subsection{Application to interferometric array optimization}

The van Cittert-Zernike theorem gives the visibility function, as measured by two point antennas, over the viewing window $\mc P$ as

\beq
\label{VCZ}
V ({\mb b}) = \int\limits_\mc P I (\mb{p}) e^{-2\pi i {\mb b} \cdot {\mb p}}\, \mathrm{d}\mb p.
\eeq

where $I (\mb{p})$ is the intensity of the radiation from angular direction $\mb{p}$, and $\mb b$ is the baseline, or displacement between the antennas, projected onto the plane orthogonal to the source propagation and divided by the wavelength observed. Assuming a small field of view, such that the object $I$ lives on the plane instead of the sphere, the visibility measurements can be approximated as

\beq
\label{DiscreteVCZ}
{\mb V} \approx (\Delta \mb{p})^{2} \sum_k {\mb I}(\mb{p}_k) e^{-2 \pi i \mb{b} \cdot \mb{p}_k}.
\eeq

In other words, $\Theta$ is a "partial" two-dimensional Fourier transform, where the array's baselines $\mb b$ dictate which Fourier coefficients are captured, i.e., where the visibility function is sampled. Since we capture a Fourier coefficient with every baseline, or pair of antennas, an array with $a$ antennas yields $\frac{a(a - 1)}{2}$ measurements in a snapshot. (Because an array cannot capture all the visibility information, neither $\Theta$ nor $\mb V$ are properly the Fourier transform matrix or the Fourier transform of $\mb I$, respectively. Unmeasured Fourier coefficients are absent in $\mb V$, and the corresponding rows are likewise absent in $\Theta$.)

Due to the role of the baselines in selecting the rows of $\Theta$, the array is critical in designing an incoherent $\Phi$. In the well-sampled case, the general consensus is that a centrally condensed or bell-shaped distribution of baselines produces more favorable near-in and far side-lobe patterns, and thus superior images, than a uniform distribution of baselines (Boone 2002; Holdaway 1996; Holdaway 1997; Kogan 1997; Woody 2001a, b). Such distributions are meant to pursue an ideal clean beam, which is both highly localized and free of lobes, e.g., a Gaussian. Array optimization depends on the imaging objectives in mind, however, and as we intend to undersample in a way that facilitates CS, minimizing MC is our objective. Incidentally, the calculation of MC in Eq. \ref{MC} is synonymous to that of an array's peak side-lobe when $\Psi$ is the pixel basis, and can be interpreted similarly. Early in the development of CS it was shown that a matrix built by uniformly and randomly selecting rows from a discrete Fourier transform obeys the RIP; thus, objects can be stably recovered from partial, uniform random Fourier measurements through $\ell_1$-norm minimization (Cand\`es et al. 2006a, b), the standard model for CS recovery. This fundamental result suggests that an array with a uniform random distribution of baselines should work well with CS when the pixel basis is used. In fact, (Cand\`es et al. 2006a) noted that their results should be of particular interest to interferometric imaging in astronomy.

Object sparsity is the other key factor.  A basis matrix $\Psi$ that results in a more coherent measurement matrix $\Phi$ may compensate by providing far sparser representations of the object than the natural pixel basis. In addition to the pixel basis, where $\Psi$ is simply the identity, the two-dimensional block discrete cosine transform (BDCT) is thus also considered. Used extensively in image compression, e.g., in JPEG, the BDCT divides an image into local blocks and decomposes each into a sum of cosines of different frequencies. In both bases we focus on the suitability of random and randomized arrays, as random sampling has often proven to be inseparable from CS theory. We compare their performances with that of the 'Y'-shaped configuration currently used by the Very Large Array (VLA). To study the most basic behavior of arrays, we focus on snapshot-mode observations and do not involve Earth rotation.

\subsection{Orthogonal Matching Pursuit}
\label{OMP}

OMP is a variant of the matching pursuit (MP) process, which searches for the sparsest reconstruction $\mb{\hat{I}}$ subject to $ \Phi \mb{I} = \mb{V}$ by iteratively projecting the data $\mb{V}$ onto greedily chosen columns of the measurement matrix $\Phi$ until a "best match" to the object $\mb{I}$ is found. We start with the assumption that the original object is $S$-sparse ($S$ is unknown), and that the data $\tbf{V}$ is therefore a linear combination of $S$ columns, not all $N$ columns, of $\Phi$. The goal is to find which subset of $S$ columns participate, as we have no idea which $S$ components of $\tbf{I}$ are non-zero. By choosing these $S$ columns, OMP can find $\mb{x}$, a vector of the non-zero components of the reconstruction $\mb{\hat{I}}$.

We first initialize the residual $\mb{r}_0 = \mb{V}$, the set of chosen column indices $\alpha_0 = \emptyset$, and the matrix of chosen columns $\Phi_0$ to the empty matrix. At each OMP finds the column that has the highest coherence with $V$, as this column likely contributes the most to the data.

\beq
\label{OMP1}
k_t = argmax_{1 \leq k \leq N} | \mb{r}_{t - 1} \cd \phi_k|
\eeq
where $\phi_k$ is the $k$-th column of $\Phi$. Ideally, this column represents the brightest remaining pixel in the object $\mb{I}$. We then add $k_t$ to the set of chosen column indices, and add $\phi_k$ to the matrix of chosen columns.

\beq
\alpha_t = \alpha_{t - 1} \cup \{k_t\}
\eeq

\beq
\Phi_t  = [ \Phi_{t - 1} \quad \phi_{k_t}]
\eeq
We can then solve the least-squares problem to update $\tbf{x}_t$

\beq
\label{OMP2}
\mb{x}_t = argmin_\mb{x} || \mb{V} - \Phi_t \mb{x} ||_2
\eeq
and use it to update the data and the residual.

\beq
\mb{V}_t = \Phi_t \mb{x}_t
\eeq

\beq
\mb{r}_t = \mb{V} - \mb{V}_t
\eeq
OMP then returns to Eq. (\ref{OMP1}) and iterates until $||\mb{r}||_2 \leq \delta||\mb{V}||_2$, where $\delta||\mb{V}||_2$ is the stopping criterion (here, $\delta = 0.01$ due to the noise parameters described in Sect. \ref{WellRes}). In essence, at each iteration OMP greedily choses a column from $\Phi$, and calculates a new residual by subtracting away from the visibility its orthogonal projection onto the span of all the columns chosen so far. Ideally, each iteration locates the brightest remaining pixel in the object $\mb{I}$, and removes the collective contribution from all the pixels chosen so far from the data. The final reconstruction $\mb{\hat{I}}$ has non-zero components only at the indices in $\alpha$, where its non-zero component at $k_t$ is the $t$-th component of $\mb{x}$. Here we use the most generic form of OMP, which obeys the process above with no other constraints on the reconstruction (in particular, we did not enforce non-negativity).

As formally shown in (Lannes et al. 1997), the traditional deconvolution algorithm CLEAN is a non-orthogonal MP process. Each iteration greedily chooses the brightest remaining pixel in the image and removes its contribution to the data, until the residual falls below some threshold. The critical difference between OMP and non-orthogonal MPs like CLEAN is Eq. (\ref{OMP2}): CLEAN does not project the data onto all the columns chosen so far, which enables OMP to flexibly update the components in the directions of past chosen columns at every iteration. The basic CLEAN simply removes the component of the data on the column chosen in the current iteration. This key difference from CLEAN makes OMP the superior recovery algorithm: when the dimension of $\mb{x}$ is large, the orthogonal projection enables the OMP residual to converge to 0 far more quickly than the MP residual. Unlike MP, OMP is also guaranteed to converge in finite iterations in finite-dimensional spaces (Davis et al. 1994; Davis et al. 1997). OMP and its many variants are widely used in CS, and their structural closeness to CLEAN is part of the natural suitability of CS for radio interferometry.

\section{Arrays}
\label{Arrays}

\begin{figure}
{\includegraphics[width=8.5cm]{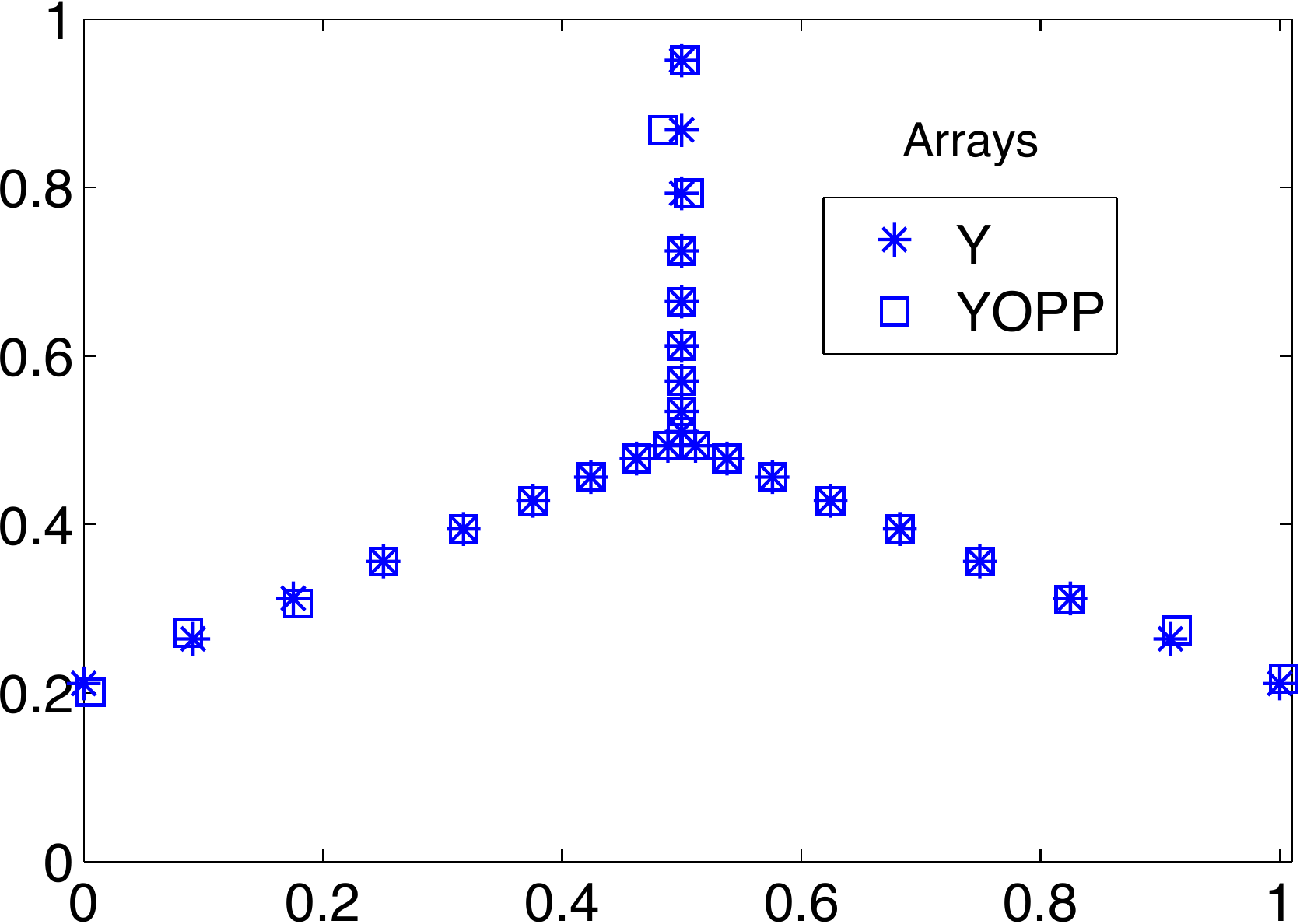} \includegraphics[width=8.5cm]{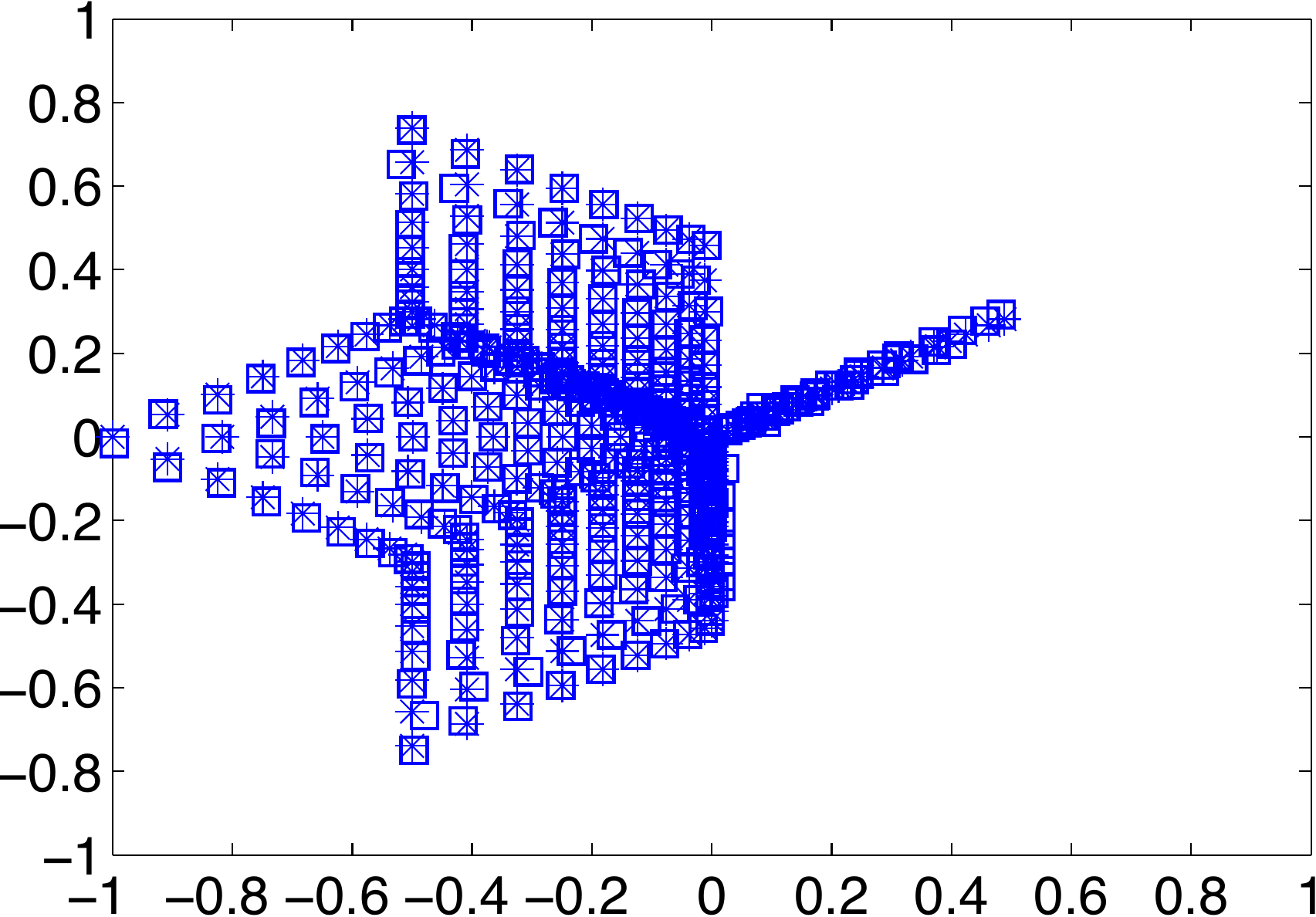}}
\caption{Antenna distribution of the modified array 'YOPP' (left) and its baseline distribution (right), compared to the antenna and baseline distribution of the VLA-based 'Y' array. Original versions of these figures appeared in (Fannjiang 2011) in SEG TLE.}
\label{YOPP}
\end{figure}

The arrays under study are the uniform random array (URA); the truncated normal random array (NRA), where sampling coverage beyond the predetermined aperture is removed; an array defined by the Hammersley point set, a low-discrepancy sequence widely used in quasi-Monte Carlo methods (Neiderreiter, 1987); and a VLA-based 'Y' configuration. (In 'Y' arrays of other than 27 antennas, the distances between antennas along an arm are based on the current ratios of distances). We also create a modification of the 'Y' array, called YOPP ('\tu{Y}', \tu{o}utermost antennas \tu{p}erpendicularly \tu{p}erturbed), in which the three outermost antennas on each arm of the 'Y' are perturbed a random distance in the direction perpendicular to their respective arm. The maximum perturbation distance is set to 30\% of the mean distance between adjacent antennas on an arm (see Fig. \ref{YOPP}). In arrays of other than 27 antennas, the outermost third of the antennas on each arm are perturbed in the same way.

The standard deviation (SD) of an NRA describes the continuous normal distribution sampled by the antenna locations (it is not calculated with respect to the actual discrete distribution of antennas). Also note that the baseline distribution, not the antenna distribution, determines where the \textit{u-v} plane is sampled. In particular, the URA does not correspond to a uniform random distribution of baselines.

\begin{figure}
{\includegraphics[width=8.5cm]{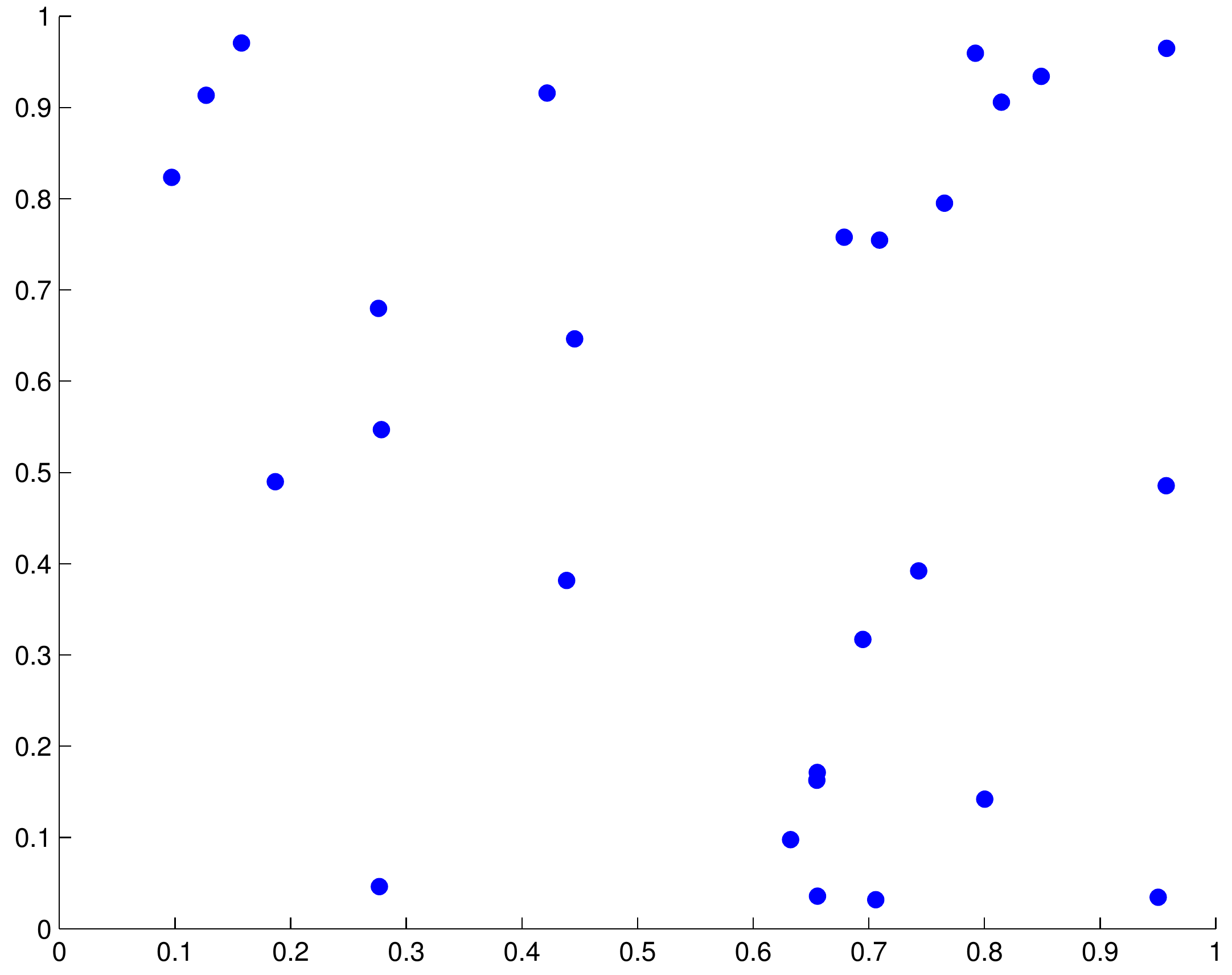} \includegraphics[width=8.5cm]{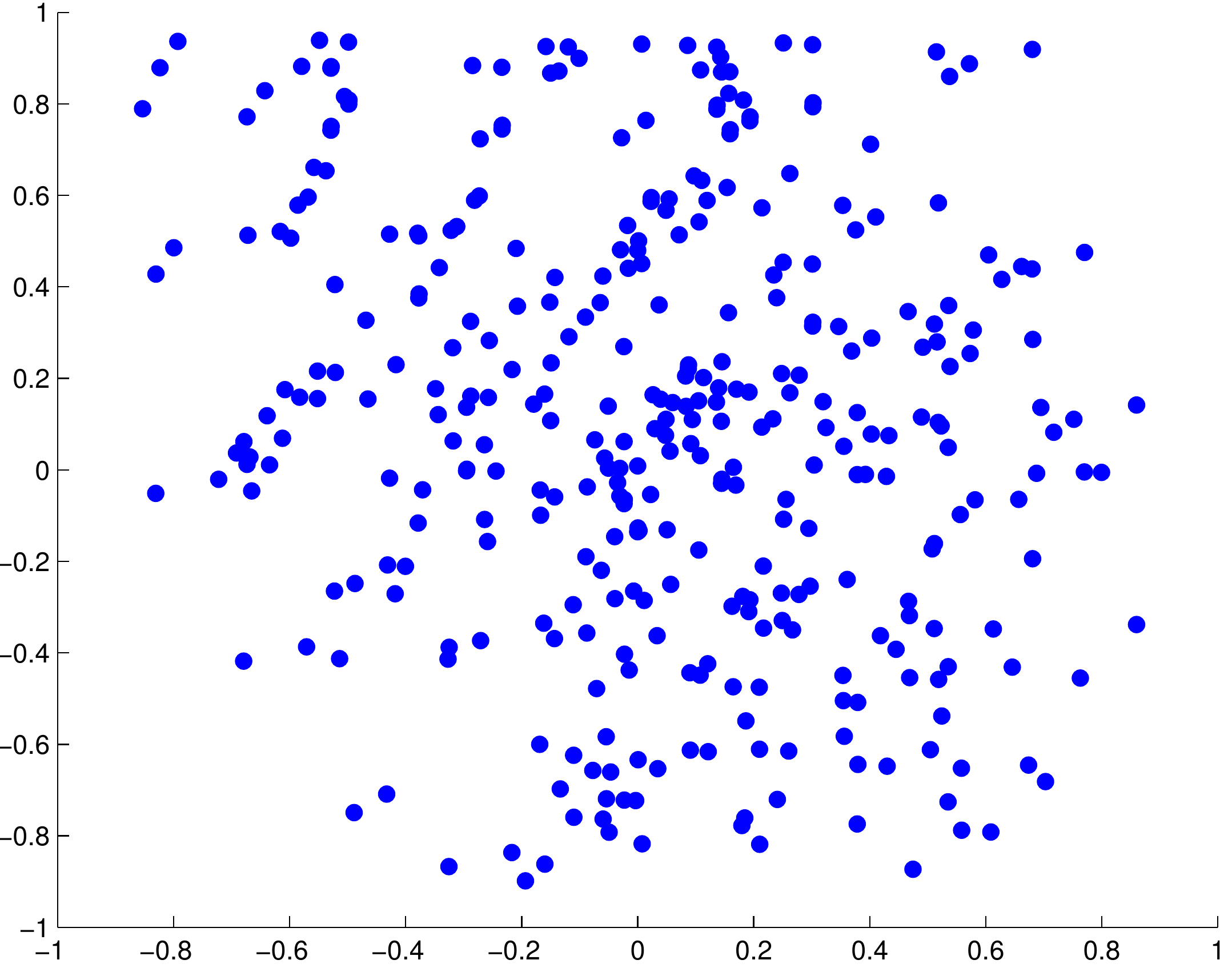}}
\caption{Antenna distribution of a URA (left) and its baseline distribution (right).}
\label{URA}
\end{figure}

\begin{figure}
{\includegraphics[width=8.5cm]{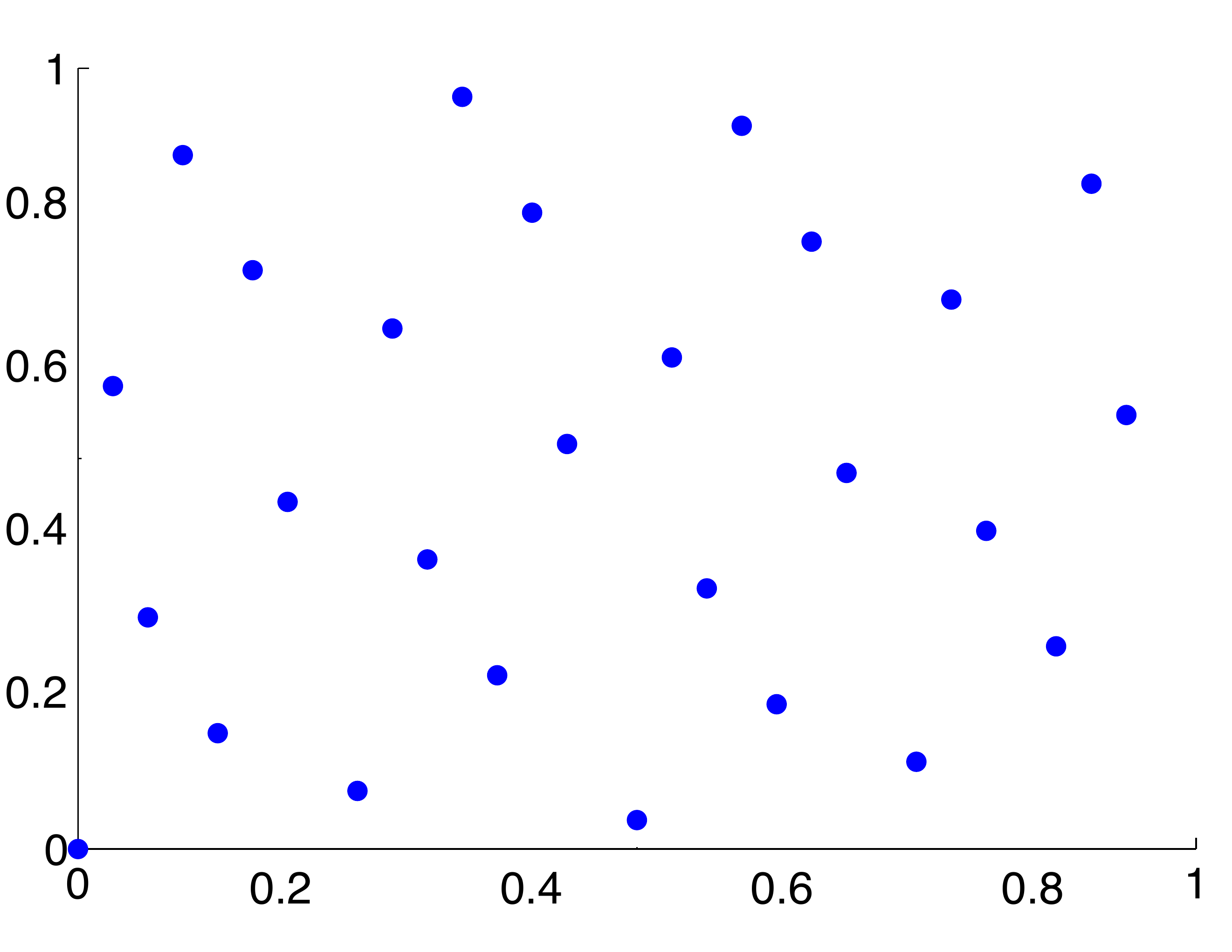} \includegraphics[width=8.5cm]{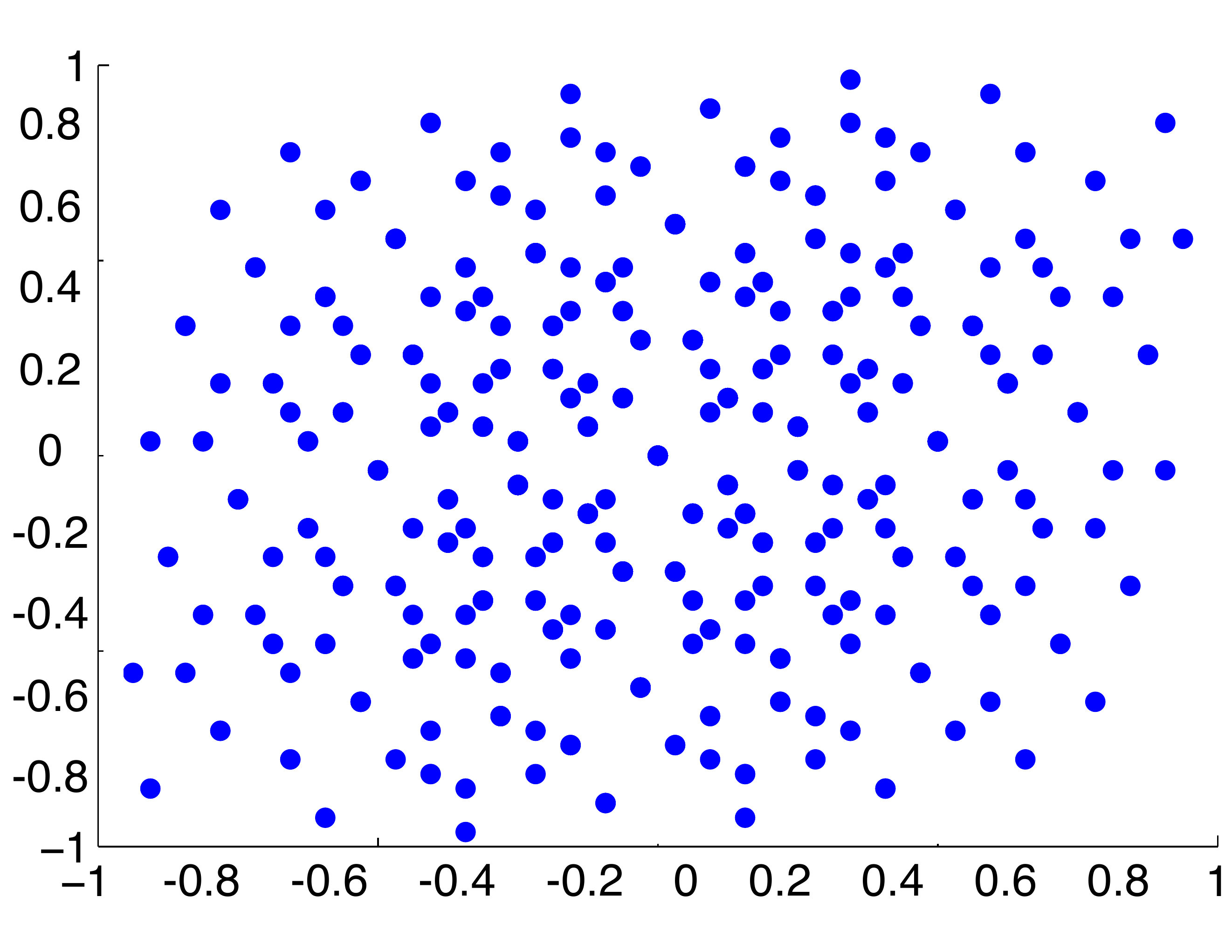}}
\caption{Antennas distributed according to the Hammersley point set (left) and the corresponding baseline distribution (right).}
\end{figure}

\begin{figure}
{\includegraphics[width=8.5cm]{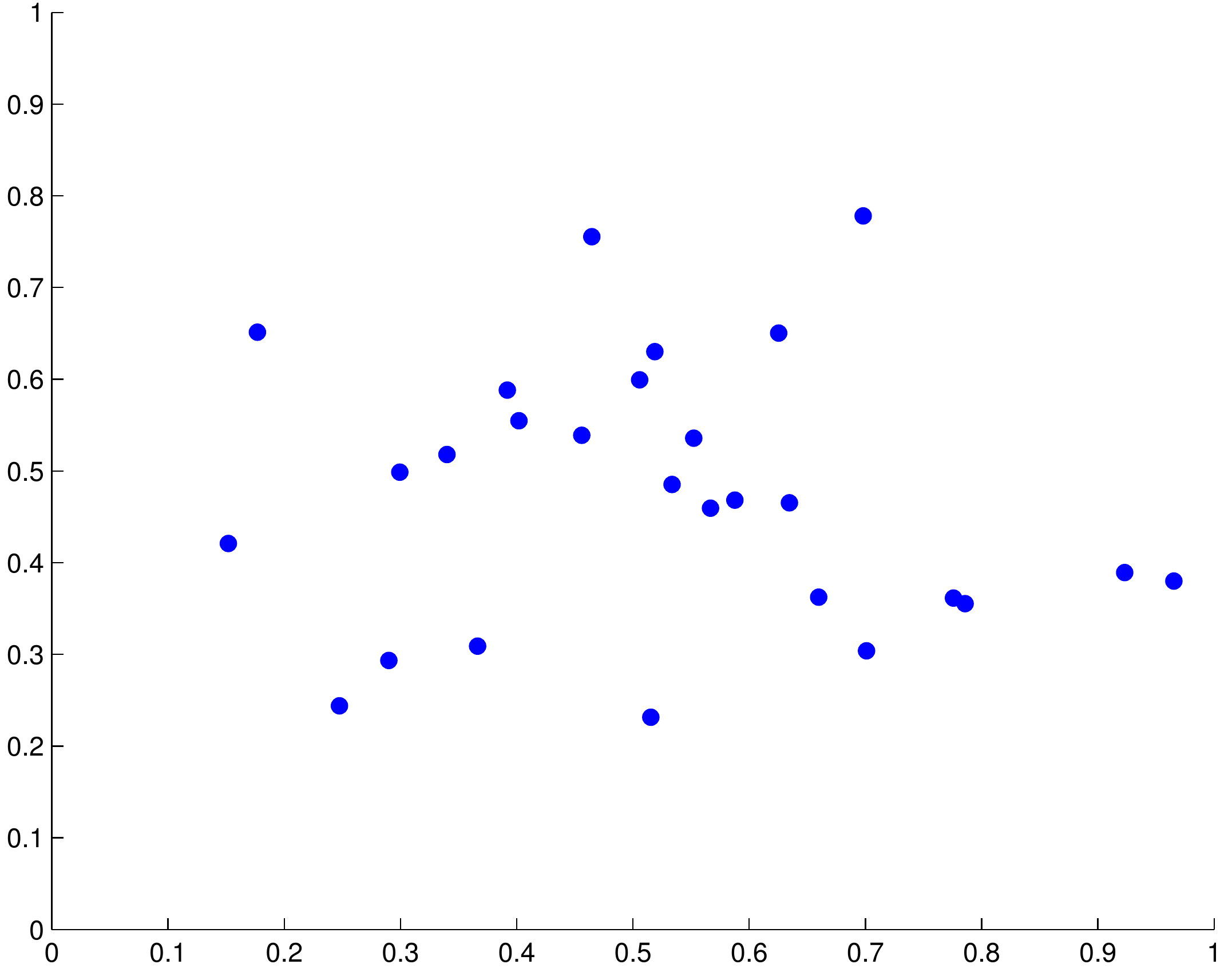} \includegraphics[width=8.5cm]{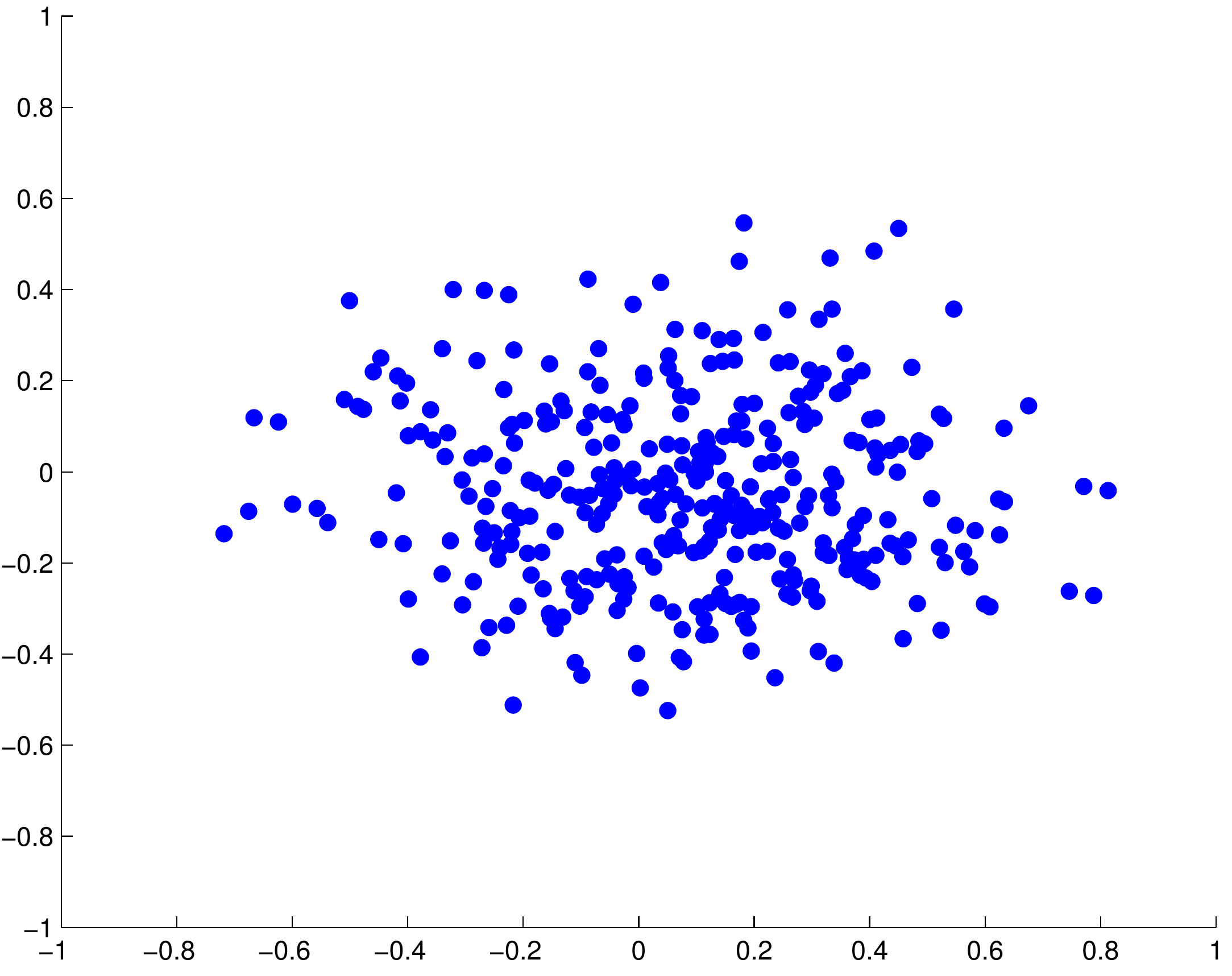}}
\caption{Antenna distribution of an NRA with SD = 0.18 (left) and its baseline distribution (right).}
\label{Hamm}
\end{figure}

\section{Pixel basis in the well-resolved case}
\label{WellRes}
We compute the MC of measurement matrices as the number of antennas increases in each array. For the randomized arrays (URA, NRA, and YOPP), the MC plotted is the mean of 50 different measurement matrices. OMP reconstructions are also run of 60 $\times$ 60 sparse objects with random point sources and Gaussian white noise. The parameters of the noise Gaussian are $\mu = 0$ and $\sigma = \frac{f ||\mb{V}||_2}{\sqrt{M}}$, where $f$ is some fraction of the visibility norm (here, 0.01) and $M$ is the dimension of $\mb{V}$. For the randomized arrays, the reconstruction rate plotted is the mean of 5 different arrays on 200 different objects.To provide a direct comparison to the existing VLA configuration, 27-antenna arrays are used in the reconstructions. Correspondingly, the measurement matrices in the MC computations have the dimensions 351 $\times$ 3600, where $351 = \frac{27(27 - 1)}{2}$. In the reconstructions of random point sources in Sects. \ref{PixText} and \ref{UnifNormText}, a reconstruction $\mb{\hat{I}}$ is deemed successful if the relative error (RE) from the original object $\mb{I}$ is less than or equal to 0.01, where the RE is defined as

\beq
\label{RE}
RE = \frac{|| \mb{\hat{I}} - \mb{I}||_2}{||\mb{I}||_2}
\eeq

\subsection{Mutual coherence and random point sources}
\label{PixText}
\begin{figure}
{\includegraphics[width=8.5cm]{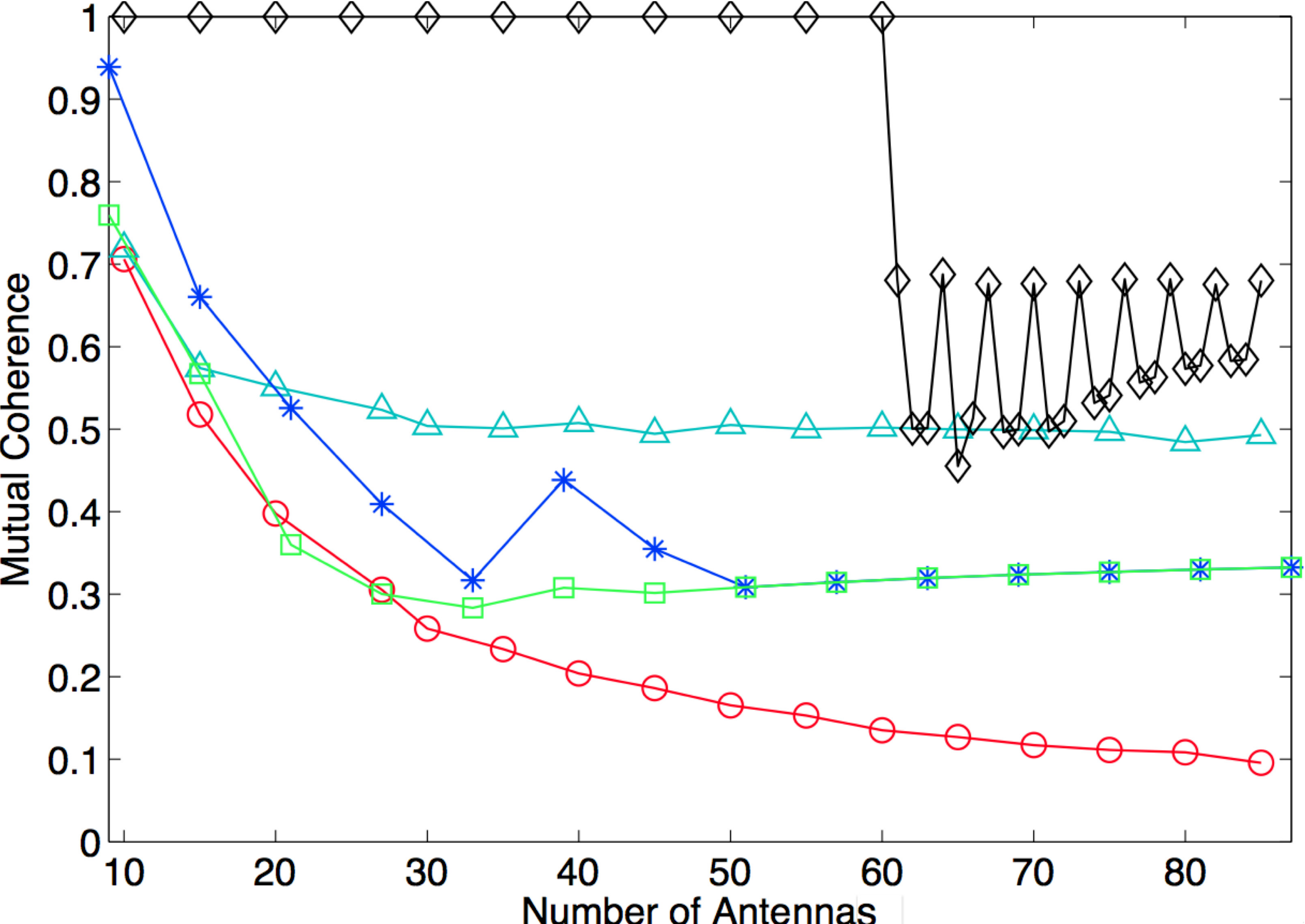} \includegraphics[width=8.5cm]{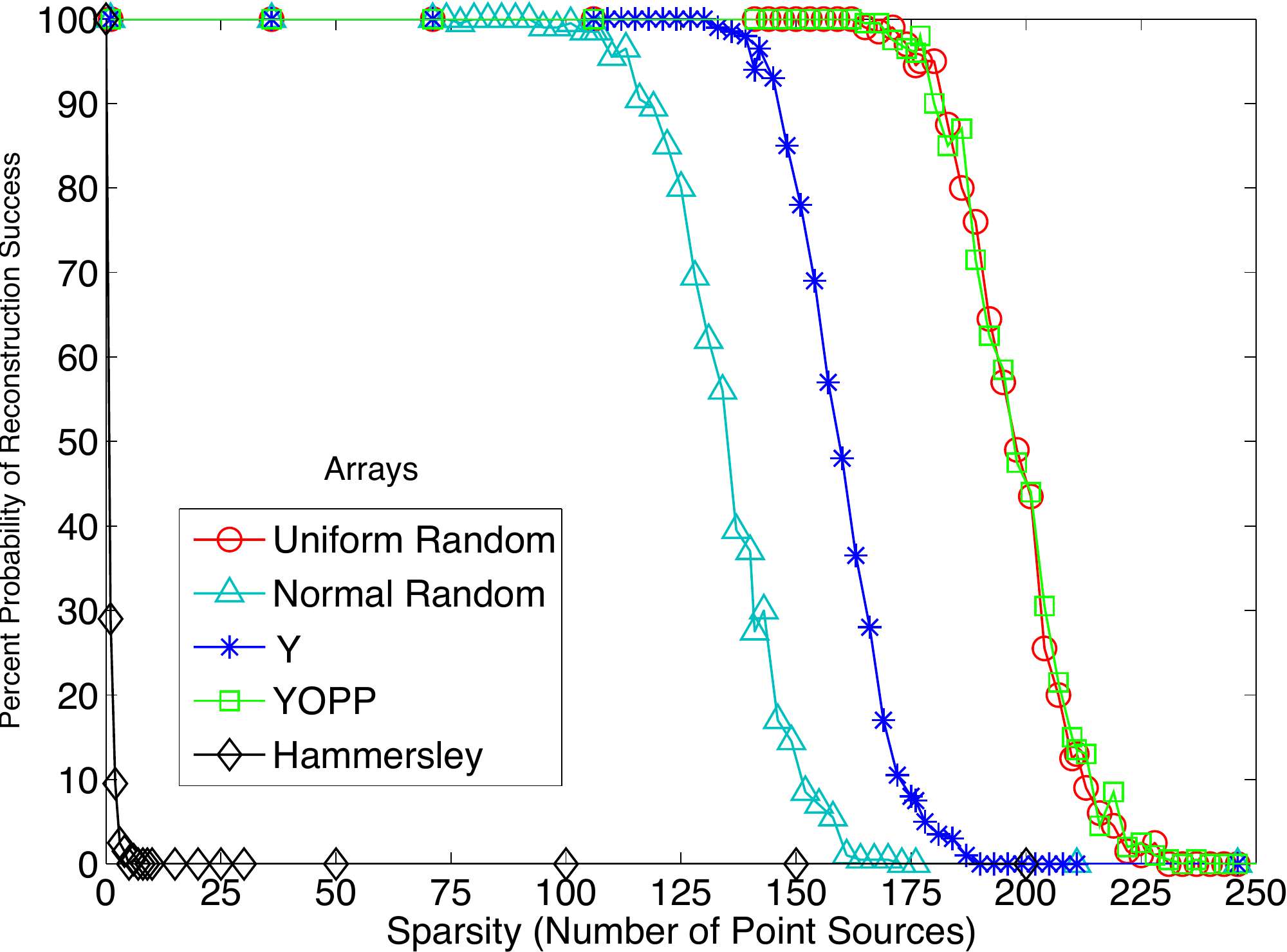}}
\caption{Mutual coherence of measurement matrices as the number of antennas in the array increases (left) and rate of successful OMP reconstructions of random point sources with arrays of 27 antennas (right). For the randomized arrays (URA, NRA, and YOPP), the MC plotted is the mean of 50 different measurement matrices, and the reconstruction rate plotted is the mean of 5 different arrays on 200 different objects. A reconstruction is deemed successful if the relative error RE $\leq$ 0.01}. Original versions of these figures appeared in (Fannjiang 2011) in SEG TLE.
\label{PixMCOMP}
\end{figure}

In Fig. \ref{PixMCOMP}, the URA provides the most incoherent measurement matrices, pointing to its suitability for OMP. It is also the only array whose MC appears to approach zero as the number of antennas increases---an important detail, as in theory it implies that OMP reconstructions can be improved indefinitely by adding more antennas to the configuration. The particular normal random array (NRA) shown does not reveal how MC varies by SD (see Sect. \ref{UnifNormText}); here, with SD = 0.14, the array provides highly coherent measurement matrices. The 'Y' array also gives a higher MC than the uniform random array, as its distinct patterns likely strengthen correlations between the columns of $\Phi$. Similarly, the Hammersley array is generated by a deterministic formula, and the resulting correlations in the measurement information likely cause its poor performance. The MC curves of both these arrays also contain aberrations; with such rigid patterns, even the slightest deviations in the relationships between antennas can throw off trends in the MC. 

The randomized modification YOPP performs nearly identically to the URA when the number of antennas is not too high, e.g. with 27 antennas. This has exciting implications for the VLA, and likely other patterned arrays as well---with only slight random perturbation to just 9 antennas, the MC is significantly decreased, tailoring it for CS application. The benefits do not appear to last as the number of antennas grows large, but for a moderate number it is a surprisingly effective improvement.

Corresponding to the MC results, the OMP reconstruction performances of the URA and YOPP with 27 antennas are nearly identical. Both provide the highest "threshold" sparsity, or maximum object sparsity with 100\% reconstruction success---about 200 point sources---and are clearly superior to the 'Y' array and NRA, which provide threshold sparsities of about 100 and 125, respectively. The high MC of the Hammersley array results in a performance that decays rapidly with only a handful of point sources.

\subsection{Uniform and normal random arrays}
\label{UnifNormText}
This section examines the spectrum from centrally condensed to uniform baseline distributions, which has been a focus of array design in the well-sampled case. Specifically, we look at normal and uniform random arrays of antennas. As the SD of the NRA decreases and the array becomes more concentrated, the measurement matrices grow more coherent; conversely, as the arrays become less concentrated, the MC decreases and approaches that of the URA. The OMP reconstruction results with 27 antennas largely follow these trends; however, there are two interesting differences. Though the MC curves decrease at a roughly linear rate as the SD of the array increases, the reconstruction probability curves do not appear to increase in any consistent mathematical relationship to the MC results, as the bounds in (Donoho et. al. 2006) would suggest. 

Secondly, though an NRA cannot better the reconstruction performance of a URA, it provides the same threshold sparsity with SD $\geq 0.16$. The measurement matrix of an NRA with an SD of 0.16, however, is still much more coherent than that of a URA. These discrepancies between the two experiments, though subtle, indicate the presence of factors at play other than MC, an observation that becomes more relevant with the BDCT in Sect. \ref{BDCT}. 

We also emphasize that while the NRA may be able to imitate the URA with 27 antennas, its MC behavior still differs fundamentally as it fails to continue approaching zero as the number of antennas grows.

\begin{figure}
{\includegraphics[width=8.5cm]{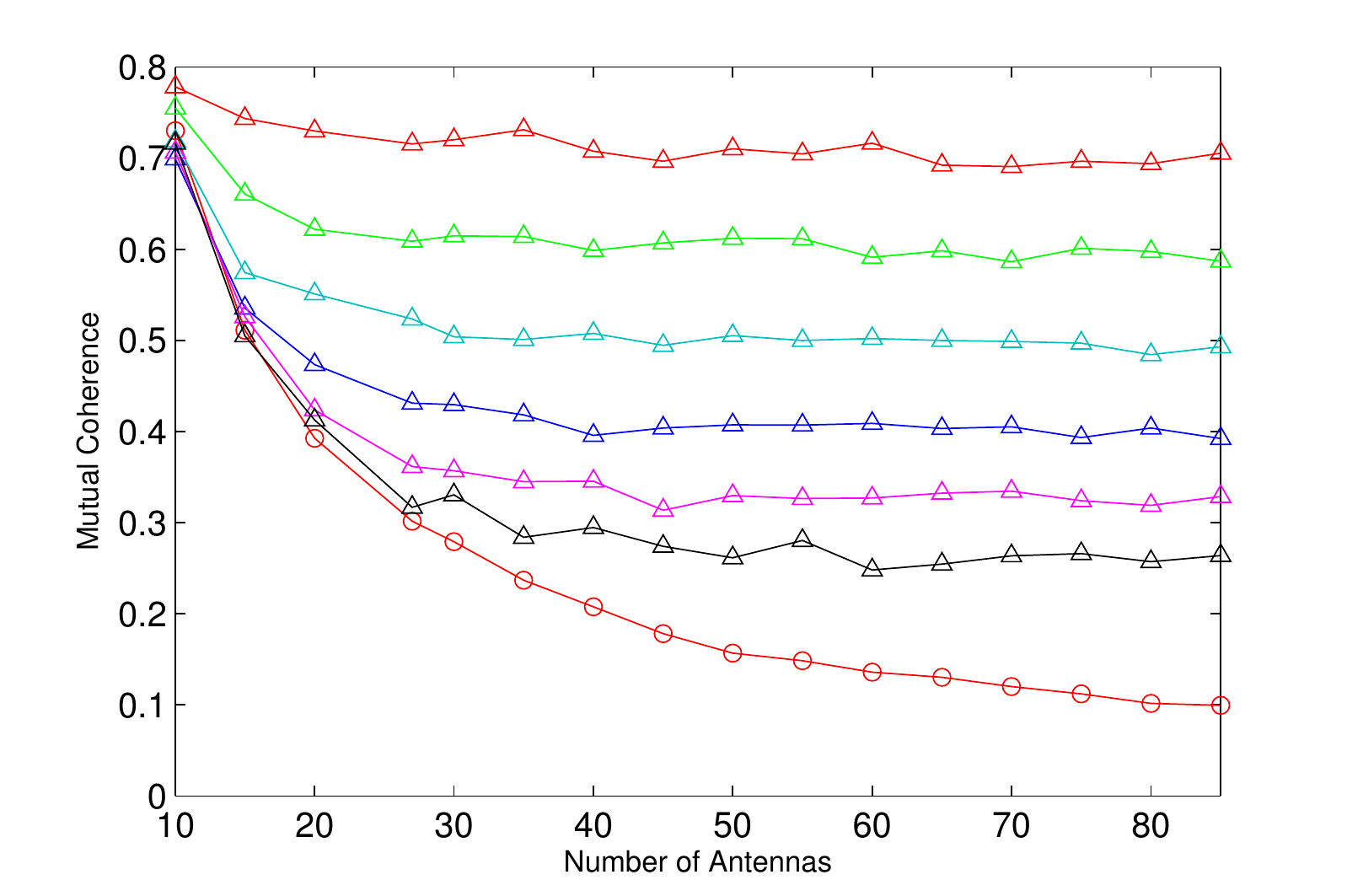} \includegraphics[width=8.5cm]{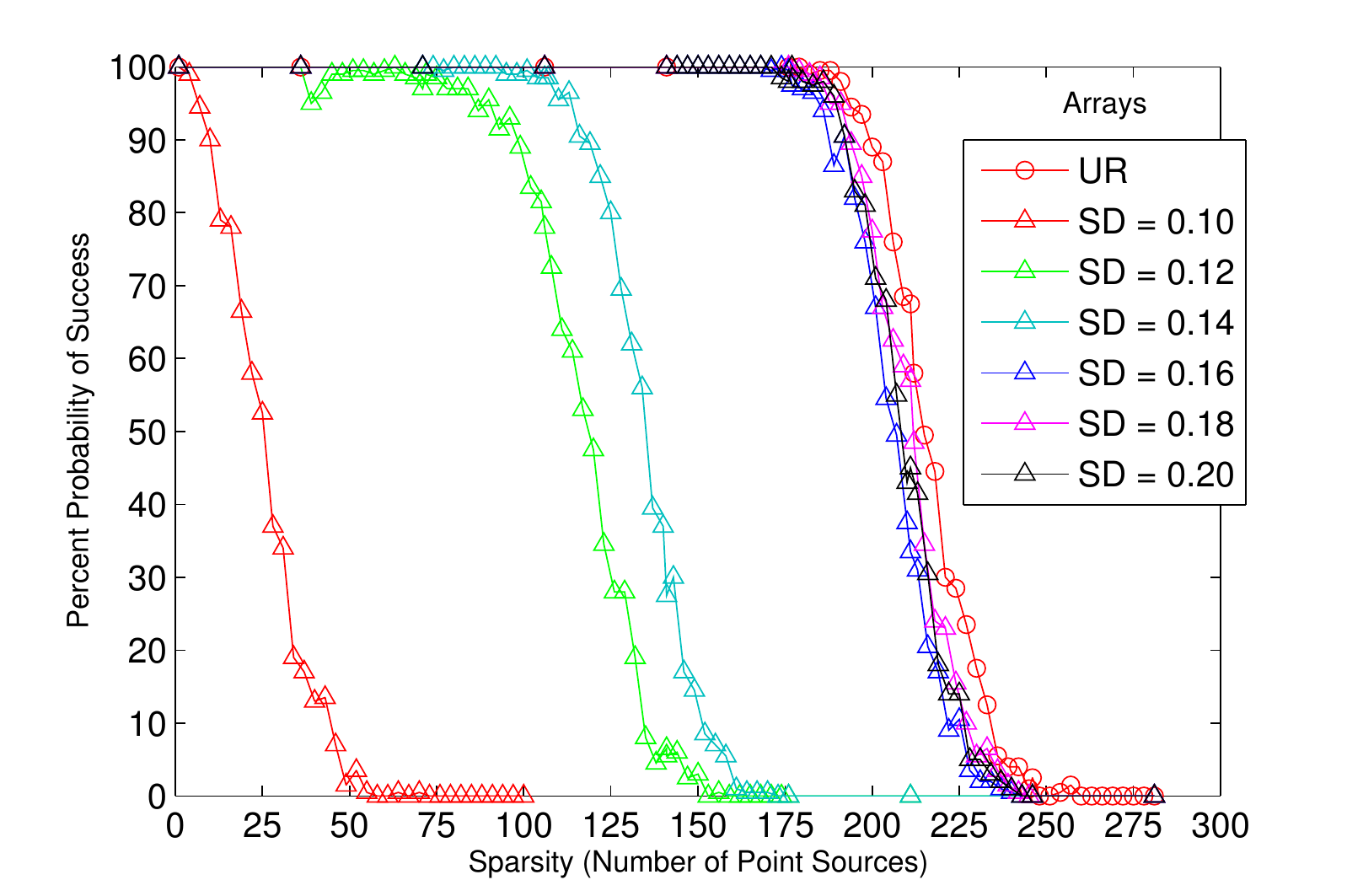}}
\caption{Mutual coherence of measurement matrices as the number of antennas in the array increases (left) and rate of successful OMP reconstructions of random point sources (right) as the standard deviation (SD) of the normal random array is increased. For the randomized arrays (URA, NRA, and YOPP), the MC plotted is the mean of 50 different measurement matrices, and the reconstruction rate plotted is the mean of 5 different arrays on 200 different objects. A reconstruction is deemed successful if the relative error RE $\leq$ 0.01}
\label{UnifNorm}
\end{figure}

\subsection{Extended object reconstructions}
\begin{figure}
\resizebox{\hsize}{!}{\includegraphics{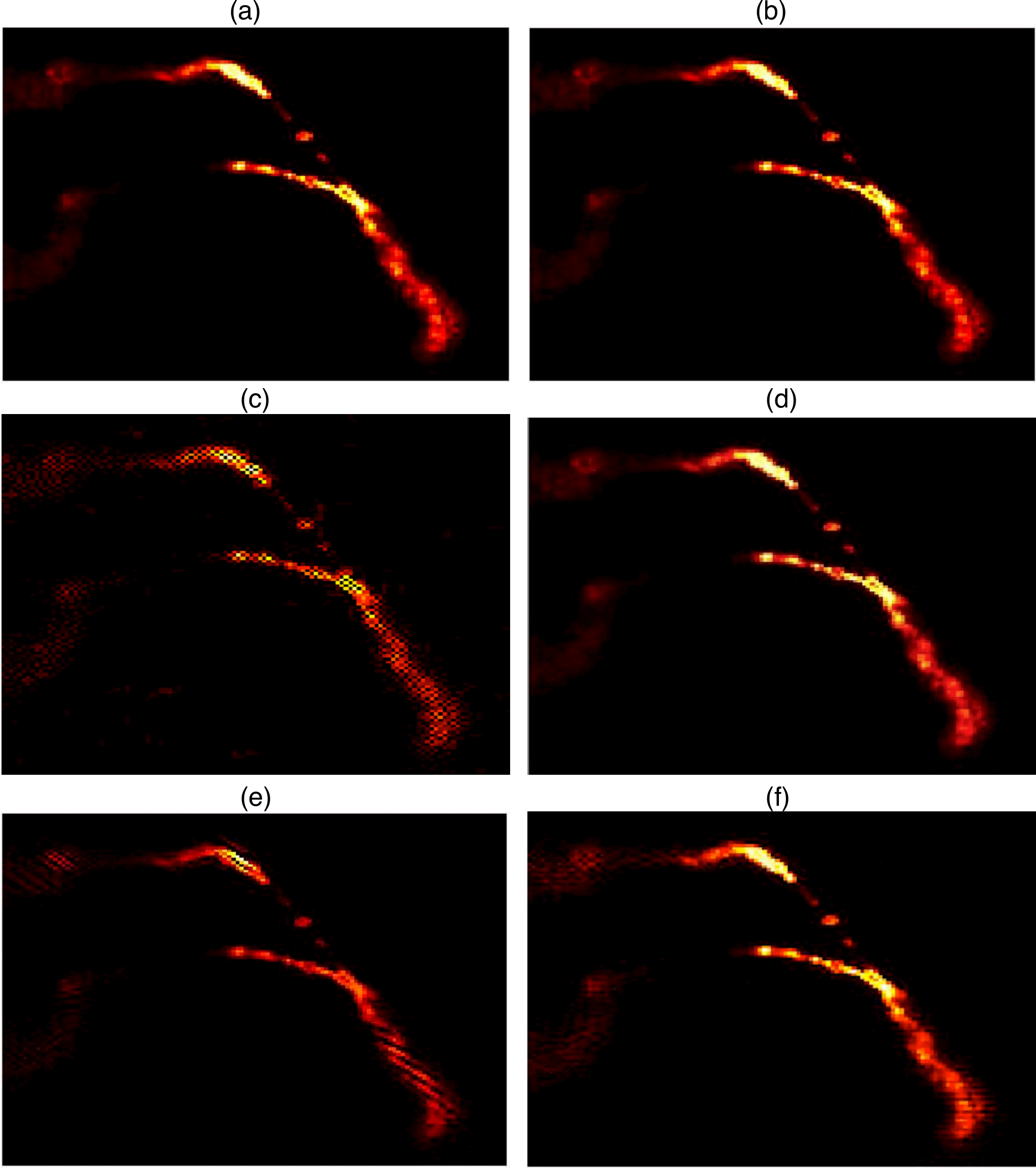}}
\caption{OMP reconstructions with the pixel basis of the object (a), emission from the radio source 3C75, simulated with a URA in (b) (RE = 1.69 $\times 10^{-4})$, an NRA of SD = 0.14 in (c) (RE = 0.763), an NRA of SD = 0.18 in (d) (RE = 0.0150), the VLA-based 'Y' array in (e) (RE = 0.179), and the YOPP array in (f) (RE = 0.0187). All pixel brightness values in the object and reconstructions are normalized by the object's maximum pixel brightness value, to lie on [0, 1]. Image (a) can be found at http://images.nrao.edu/29 (courtesy of NRAO/AUI and F.N. Owen, C.P. O'Dea, M. Inoue, and J. Eilek). Original versions of (b), (c), (e), and (f) appeared in (Fannjiang 2011) in SEG TLE.}
\label{Pix3C75}
\end{figure}

The probability curves in the previous subsections were based on reconstructions of random point sources. Here we broaden those results to reconstructions of a more realistic extended source, the black hole system \object{3C75}.  A 120 $\times$ 120-pixel image is used as the object, where all pixel brightness values are normalized by the maximum brightness value to lie on [0, 1]. Snapshot-mode data are simulated from 100 antennas. Again, the URA produces an exceptionally accurate reconstruction of the object in Fig. \ref{Pix3C75} (b). The performances of the two NRAs reflect the nuances in antenna concentration seen in Fig. \ref{UnifNorm}; though the array of SD = 0.14 produces a distinct checkerboard pattern of missed pixels, increasing the SD to just 0.18 erases these artifacts and greatly improves the accuracy of the reconstruction. Visually speaking, its reconstruction appears comparable to the URA's benchmark. 

The checkerboard artifact displayed by the NRA of SD = 0.14 is an interesting phenomenon that can be linked to high MC when pixels of non-zero intensity are adjacent (rather than randomly scattered, as in the previous subsections). Adjacent columns of $\Phi$, corresponding to the measurement information of adjacent pixels, naturally tend to be more coherent than columns that are far apart. Thus, OMP is more likely to misinterpret which of two given columns contributed most to the data (equivalently, which of the two pixels is brighter in the object) if they are adjacent. This mishandling of adjacent pixels is more prone to occur if $\Phi$ has a high MC, and could thus result in the checkerboard pattern as well as the striped artifacts produced by the 'Y' array. The slight randomization in the YOPP array remedies these latter artifacts enormously, again proving to be a simple yet effective modification. Though the noise-like left half does displays faint checkerboard artifacts, the main body of the galaxy is reconstructed accurately and the relative error (RE) is comparable to that of the NRA of SD = 0.18.

The point and extended source reconstructions both confirm the URA as the ideal array for CS with the pixel basis, notably over patterned arrays like the VLA. This finding counters that of (Wenger et al. 2010), the only other study on interferometric arrays for CS to our knowledge, which found that a uniform random baseline distribution (URB) performs significantly worse than the patterned VLA baseline distribution. In fact, any randomization added to the VLAB, similar in spirit to the YOPP presented here, degraded the quality of the CS reconstruction. These results do not align with established principles of CS theory. A measurement matrix $\Phi$ generated with a URB is precisely the matrix of uniform, random Fourier measurement vectors proven in (Cand\`es et al. 2006a, b) to obey conditions guaranteeing stable CS recovery. Though we used a URA, whose baselines do not strictly follow a uniform random distribution, the underlying principles of uniformity and randomness make our results much more plausible under a theoretical light. Random, as opposed to deterministic, sampling has always been a basic tenet of CS theory.

\section{Super-resolution with the pixel basis}
\label{SuperRes}
Enhancing angular resolution, while maintaining the integrity of the information collected, is always of interest in astronomical imaging. In this section, we numerically test the suitability of arrays for CS in the under-resolved case. The diffraction limit is broken by shrinking the aperture of the array such that $\frac{AR}{\lambda} < 1$, where $A$ is the aperture or the longest baseline, $R$ is the angular resolution, and $\lambda$ is the signal wavelength. All other experimental parameters follow those described in Sect. \ref{WellRes}.

\subsection{Array comparisons with random point sources}
For arrays of 27 antennas, the URA again provides the lowest mutual coherence, and by a far greater margin than in the well-resolved case when $\frac{AR}{\lambda}$ falls below 0.8. The OMP reconstruction curves, simulated when $\frac{AR}{\lambda} = 0.5$, confirm the URAÕs capacity for super-resolution of point sources. While it maintains a 100\% success rate beyond objects of 100 point sources, no other tested array displays anything near this potential. The URA's capabilities are likely due to its higher proportion of large baselines, which results in an increased sensitivity to the high-frequency Fourier components of an object. The array's reconstructions of point sources in Sect. \ref{WellRes} are probably superior for the same reason. 

As decreasing aperture size has the effect of condensing an array, it is intuitive to reason that a far greater SD than in the well-resolved case is needed to imitate a URA. Similarly, the random perturbation in YOPP does not appear to improve the patterns of the 'Y' array here. The maximum perturbation distance in YOPP is set based on the mean distance between antennas, and this mean distance grows irrelevantly small with smaller apertures. Far greater perturbation, perhaps to an extent that the deterministic 'Y' framework is unrecognizable, is probably needed to compensate.

\begin{figure}
{\includegraphics[width=8.5cm]{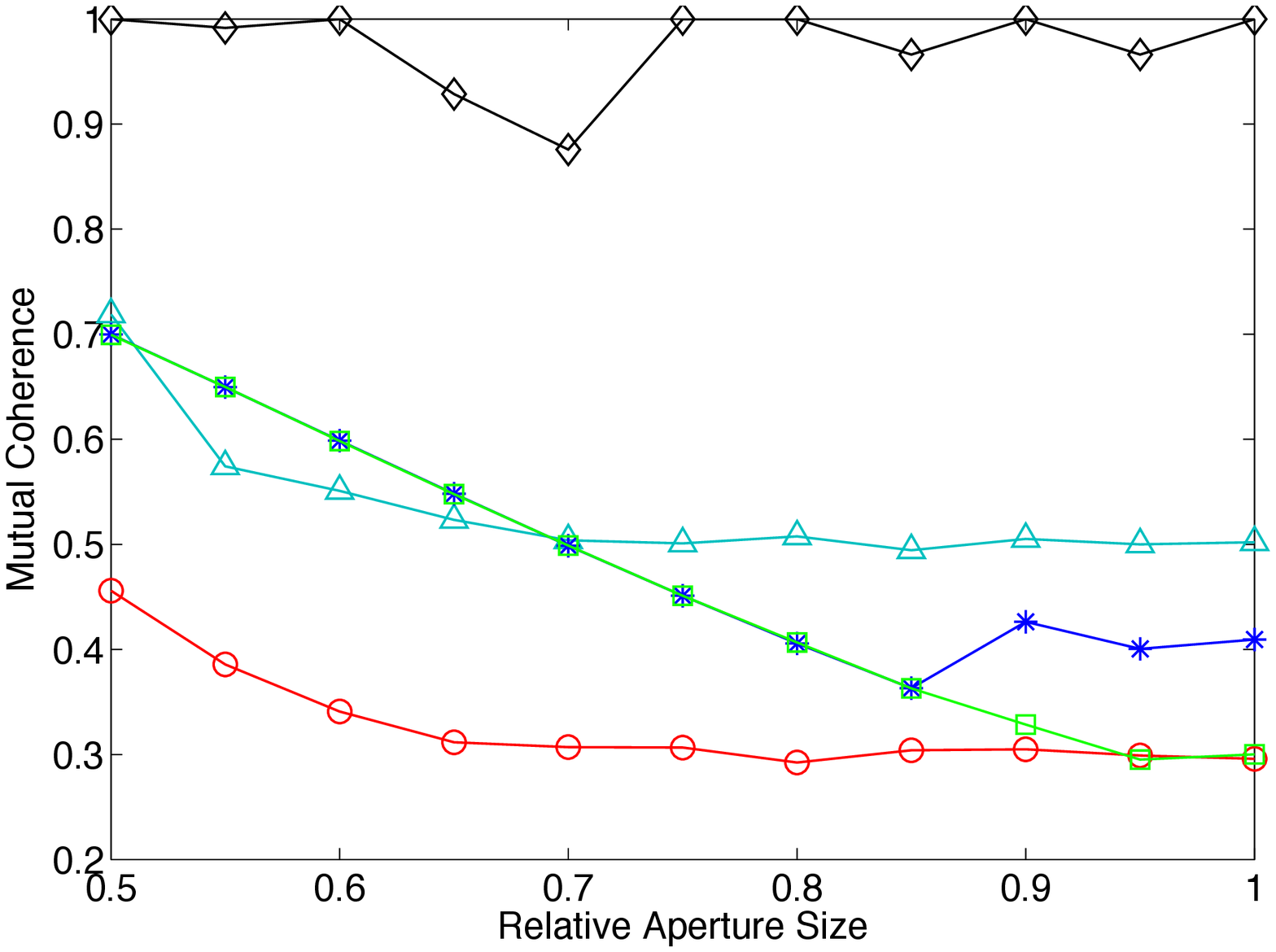} \includegraphics[width=8.5cm]{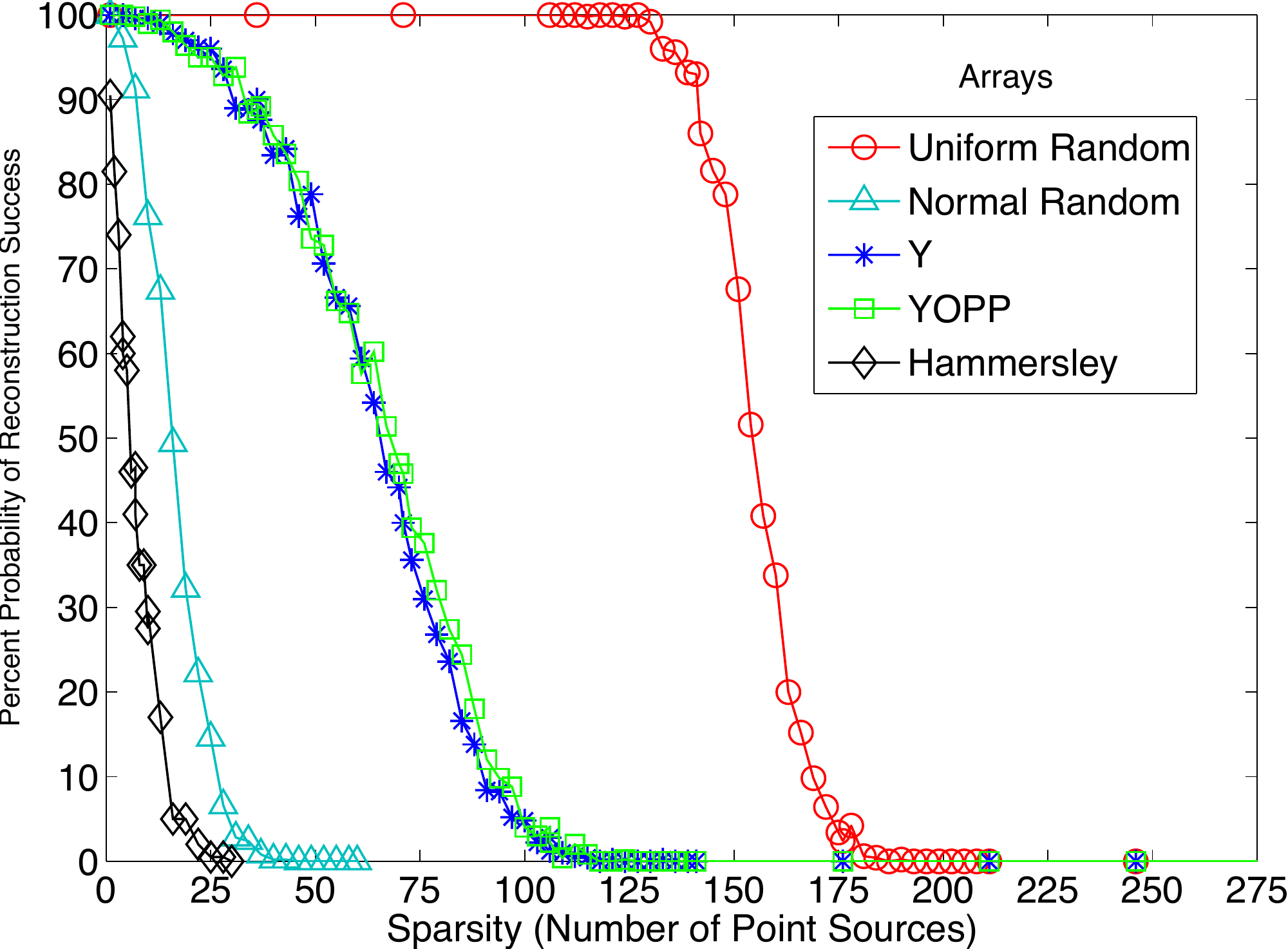}}
\caption{Mutual coherence of measurement matrices as the number of antennas in the array increases (left) and rate of successful OMP reconstructions of random point sources (right) in the under-resolved case. For the randomized arrays (URA, NRA, and YOPP), the MC plotted is the mean of 50 different measurement matrices, and the reconstruction rate plotted is the mean of 5 different arrays on 500 different objects. A reconstruction is deemed successful if the relative error RE $\leq$ 0.01. In the left panel, the aperture size is shrunk relative to the aperture size in the well-resolved case, which is set to 1. In the right panel, the relative aperture size is 0.5. Original versions of these figures appeared in (Fannjiang 2011) in SEG TLE.}
\label{PixUnder}
\end{figure}

\subsection{Extended source reconstructions}
OMP reconstructions are run of the object \object{3C75} in the under-resolved case, setting $\frac{AR}{\lambda} = 0.75$. The advantages of the URA are even more extraordinary here, as the accuracy of its reconstruction appears unaffected from the well-resolved case. Predictably, the other two arrays give much weaker performances; the 'Y' array barely outlines the galaxyÕs existence, while the NRAÕs reconstruction (using an SD that emulates a URA in the well-resolved case) is marred by artifacts similar to the checkerboard pattern seen in the well-resolved case.

\begin{figure}
\resizebox{\hsize}{!}{\includegraphics{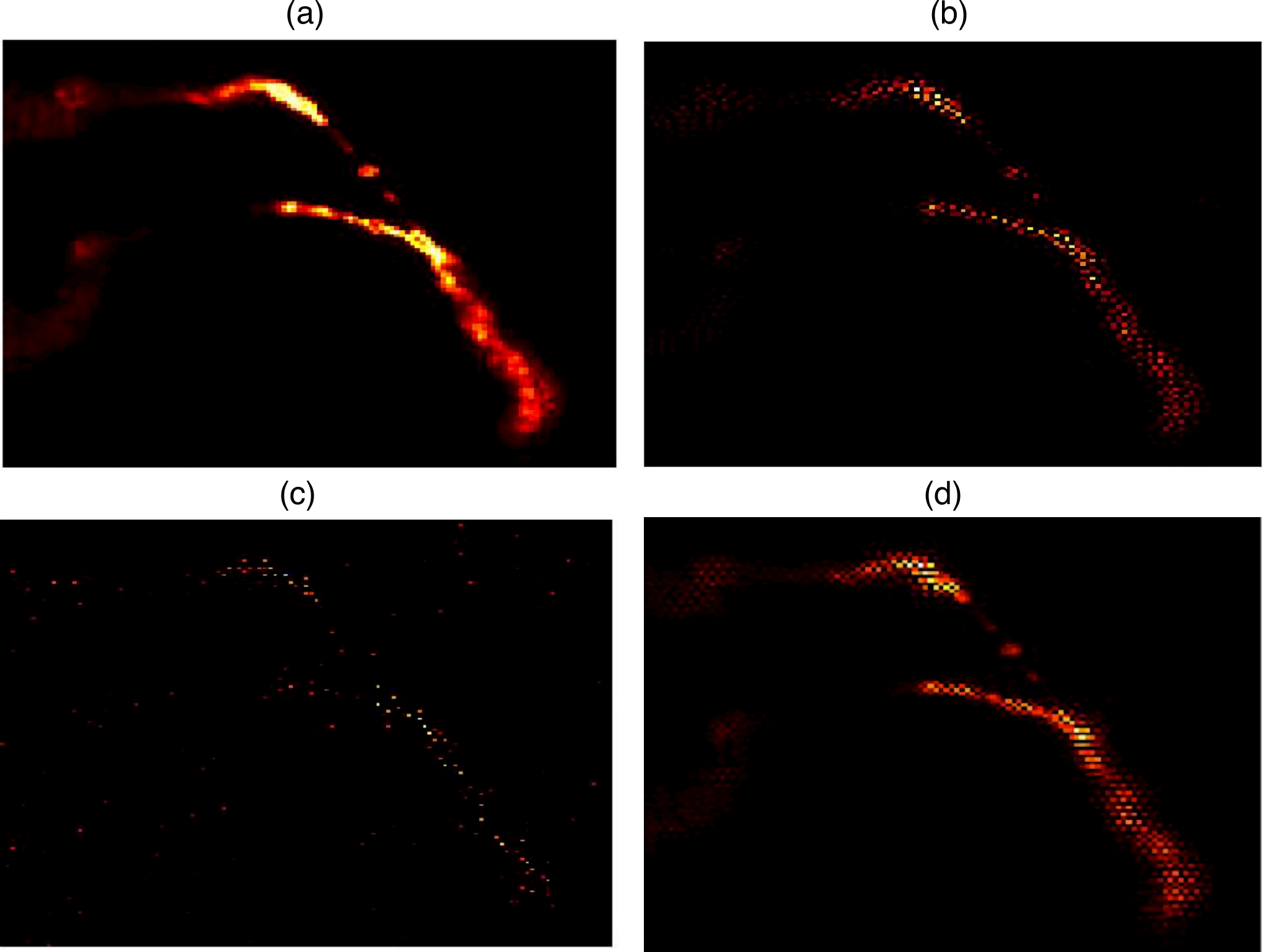}}
\caption{OMP reconstructions with the pixel basis of radio emission from 3C75 (shown in Fig. \ref{Pix3C75} (a)) simulated with a URA in (a) (RE = 2.89 $\times 10^{-4}$), an NRA of SD = 0.18 in (b) (RE = 5.4617), the 'Y' array in (c) (RE = 0.840), and the YOPP array in (d) in the under-resolved case when $\frac{AR}{\lambda} = 0.75$. All pixel brightness values in the object and reconstructions are normalized by the object's maximum pixel brightness value, to lie on [0, 1]. Image (a) can be found at http://images.nrao.edu/29 (courtesy of NRAO/AUI and F.N. Owen, C.P. O'Dea, M. Inoue, and J. Eilek). Original versions of (a), (b), and (c) appeared in (Fannjiang 2011) in SEG TLE.}
\label{Pix3C75Under}
\end{figure}

\section{The BDCT}
\label{BDCT}
Here we define the basis matrix $\Psi$ as the BDCT matrix for 4 $\times$ 4-pixel transform blocks, and $\Phi$ gave measurements for a 32 $\times$ 32-pixel object. (Unpublished results with 2 $\times$ 2- and 8 $\times$ 8-pixel blocks show the same relationships between arrays). 

\subsection{Array comparisons through mutual coherence}
\begin{figure}
\resizebox{\hsize}{!}{\includegraphics{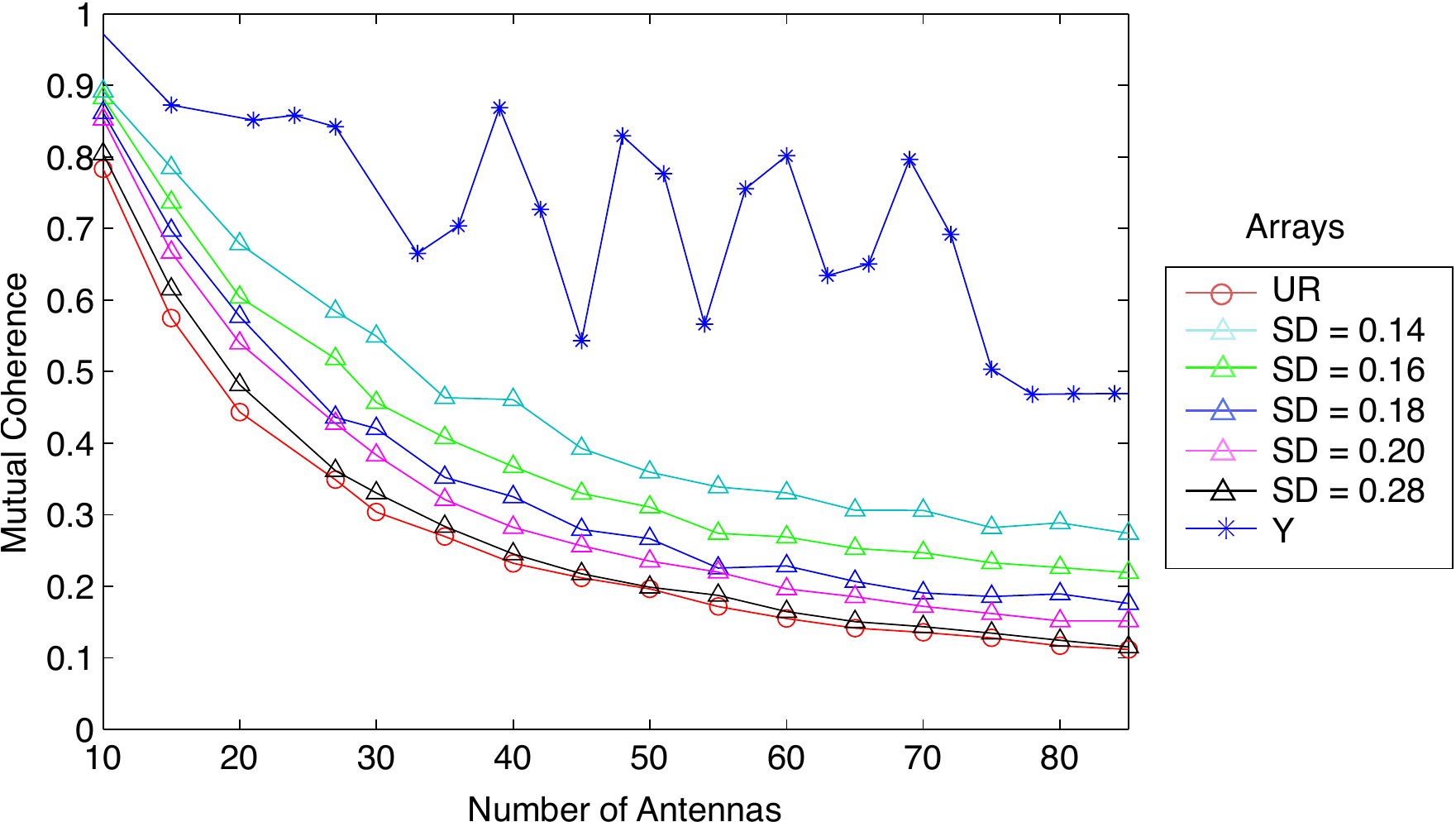}}
\caption{Mutual coherence of measurement matrices as the number of antennas in the array increases, using the BDCT matrix as the sparsity matrix. For the randomized arrays (URA, NRA), the MC plotted is the mean of 50 different measurement matrices.} 
\label{BDCTMC}
\end{figure}

Fig. \ref{BDCTMC} reveals some of the same trends as in the pixel basis; the URA provides the most incoherent measurement matrices, while the 'Y' array results in a much higher MC. Use of the BDCT also seems to have amplified the effects of patterning, as shown in the arrayÕs erratic MC curve. However, the NRAs have become more congruent with the uniform random array, for the subtle but critical detail that they decay in a like manner as the number of antennas increases. CS reconstructions in theory could thus be improved in a predictable fashion simply by adding more antennas to either array. This similarity allows an NRA of SD $>$ 0.28 to emulate the URA not just up to 27 antennas, but regardless of the number of samples. 

Though the basic pattern of higher SD resulting in lower MC is also exhibited, what is startling is how OMP reconstructions with the BDCT fail to correspond to the MC results, as seen in the next subsection.

\subsection{Reconstructing extended sources}
\begin{figure}
\resizebox{\hsize}{!}{\includegraphics{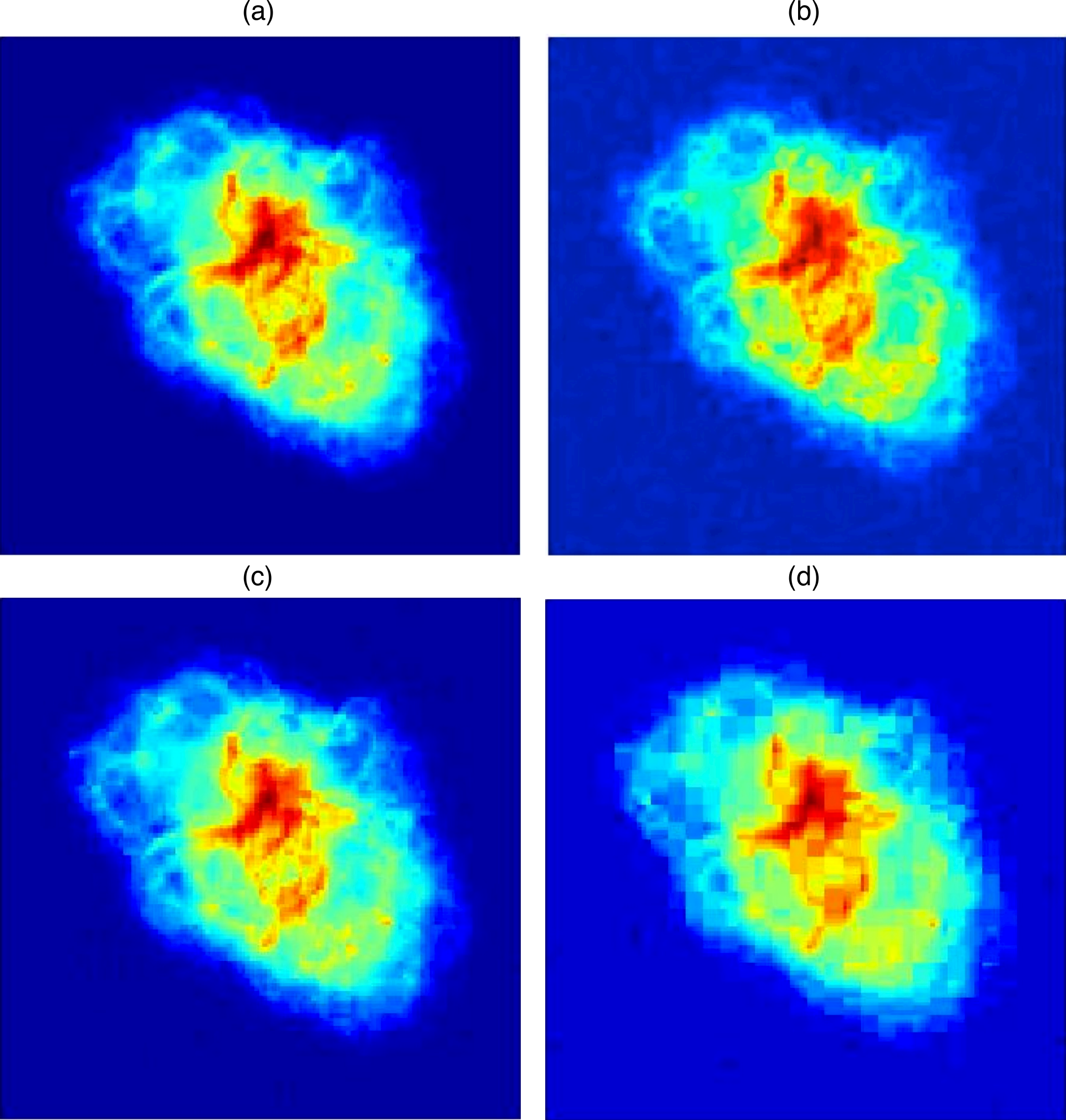}}
\caption{OMP reconstructions of the object (a), the radio emission from the \object{Crab Nebula}, using the BDCT matrix as the sparsifying matrix. The original image is of size 120 $\times$ 120 pixels and measurements are collected from 100 antennas. Reconstructions are simulated with a URA in (b) (RE = 0.0633), an NRA of 0.14 in (c) (RE = 0.0378), and the VLA-based 'Y' array in (d) (RE = 0.0752). All pixel brightness values in the object and reconstructions are normalized by the object's maximum pixel brightness value, to lie on [0, 1]. Image (a) can be found at http://images.nrao.edu/393 (courtesy of NRAO/AUI and M. Bietenholz).}
\label{BDCTCrabNeb}
\end{figure}

\begin{figure}
\resizebox{\hsize}{!}{\includegraphics{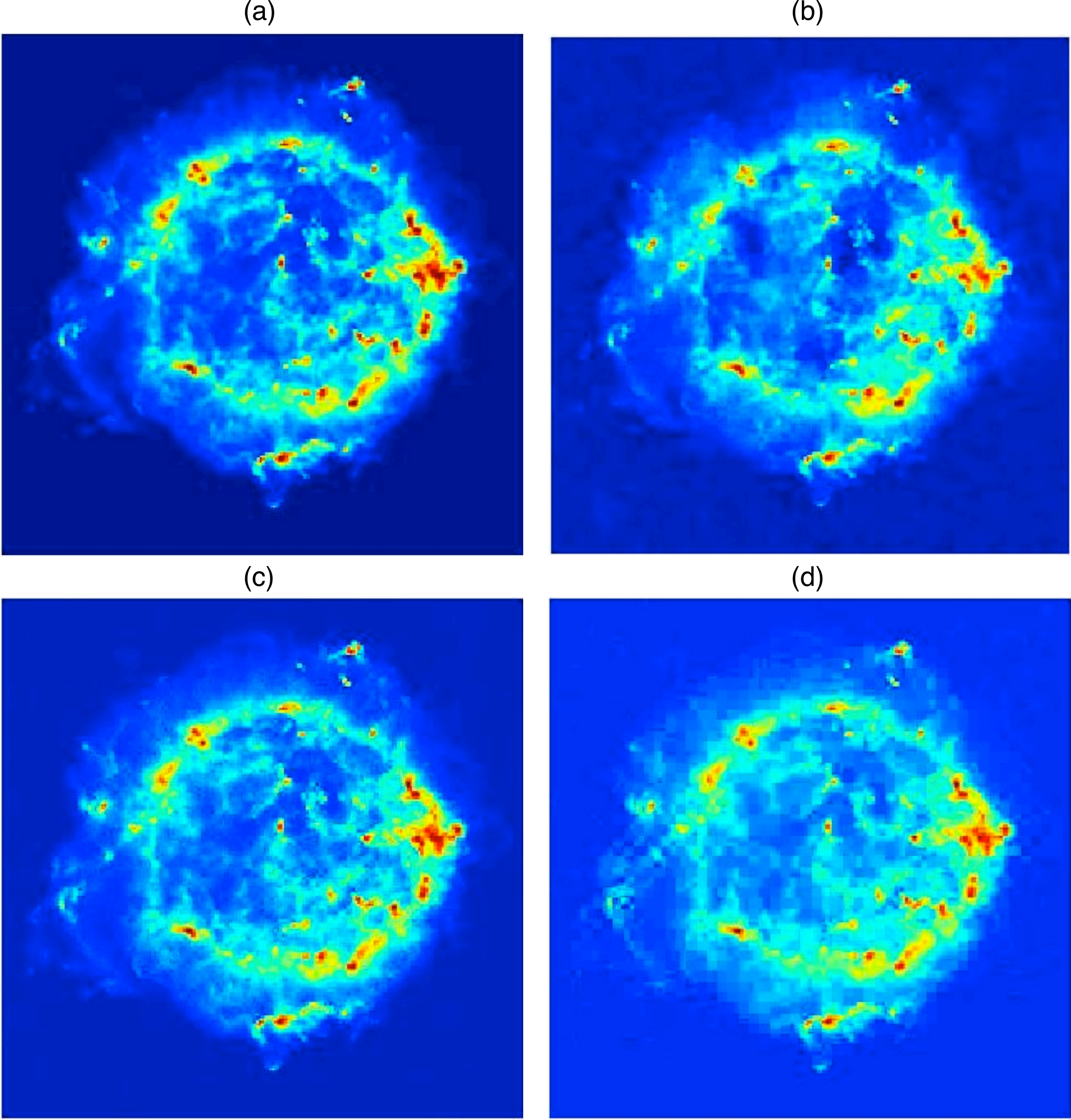}}
\caption{OMP reconstructions of the object (a), the radio emission from \object{Cassiopeia A}, using the BDCT matrix as the sparsifying matrix. The object is of size 200 $\times$ 200 and measurements are collected from 180 antennas. Reconstructions are simulated with a URA in (b) (0.0434), an NRA of 0.14 in (c) (RE = 0.0033), and the VLA-based 'Y' array in (d) (RE = 0.0119). All pixel brightness values in the object and reconstructions are normalized by the object's maximum pixel brightness value, to lie on [0, 1]. Image (a) can be found at http://images.nrao.edu/395 (courtesy of NRAO/AUI). }
\label{BDCTCassa}
\end{figure}

\begin{figure}
\resizebox{\hsize}{!}{\includegraphics{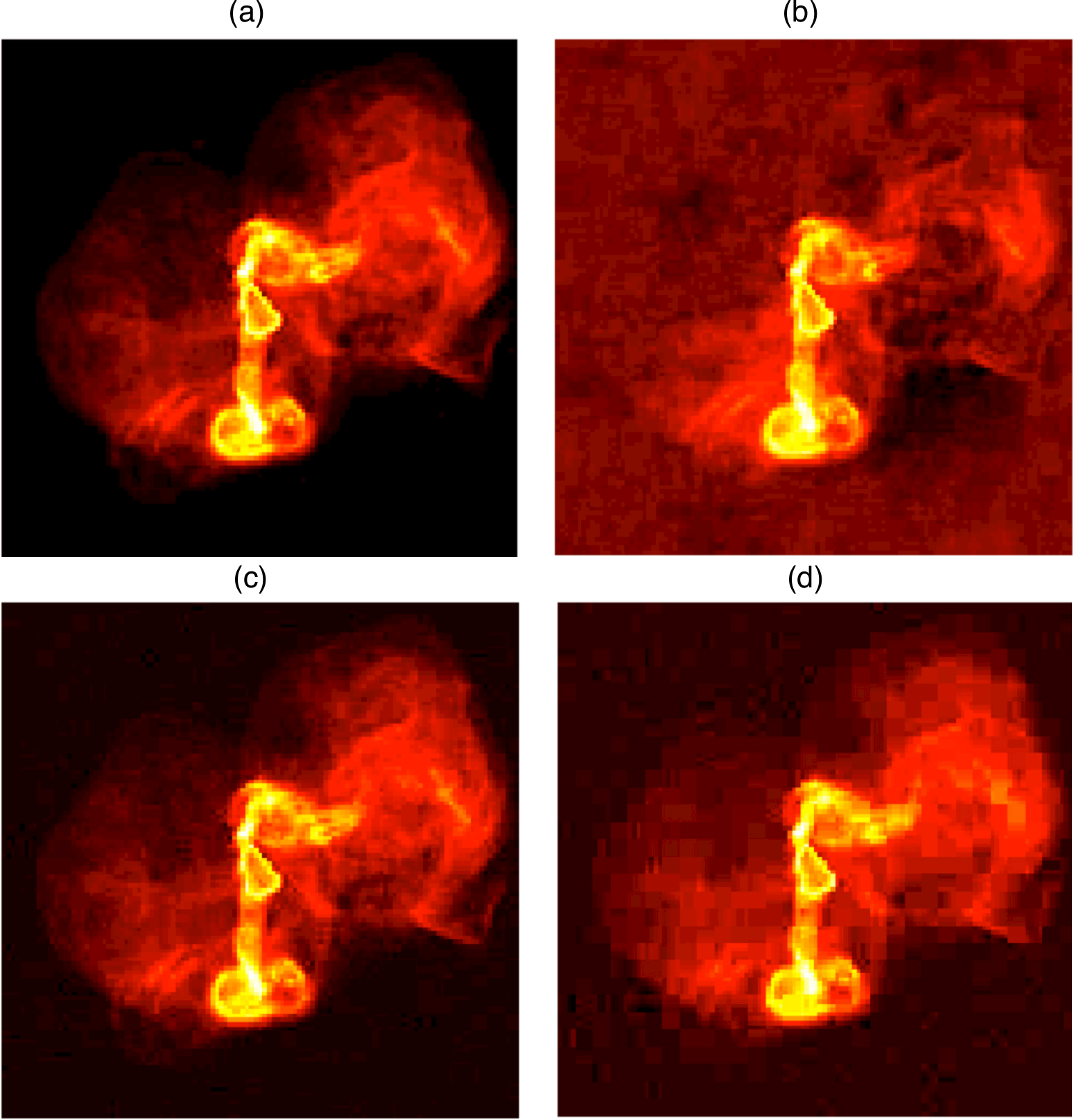}}
\caption{OMP reconstructions of the object (a), the radio emission from \object{M87}, using the BDCT matrix as the sparsifying matrix. The object is of size 120 $\times$ 120 and measurements are collected from 100 antennas. Reconstructions are simulated with a URA in (b) (RE = 0.2876), an NRA of 0.14 in (c) (RE = 0.002), and the VLA-based 'Y' array in (d) (RE = 0.009). All pixel brightness values in the object and reconstructions are normalized by the object's maximum pixel brightness value, to lie on [0, 1]. Image (a) can be found at http://images.nrao.edu/271 (courtesy of NRAO/AUI and F.N. Owen, J.A. Eilek, and N.E. Kassim).}
\label{BDCTM87}
\end{figure}

\begin{figure}
\resizebox{\hsize}{!}{\includegraphics{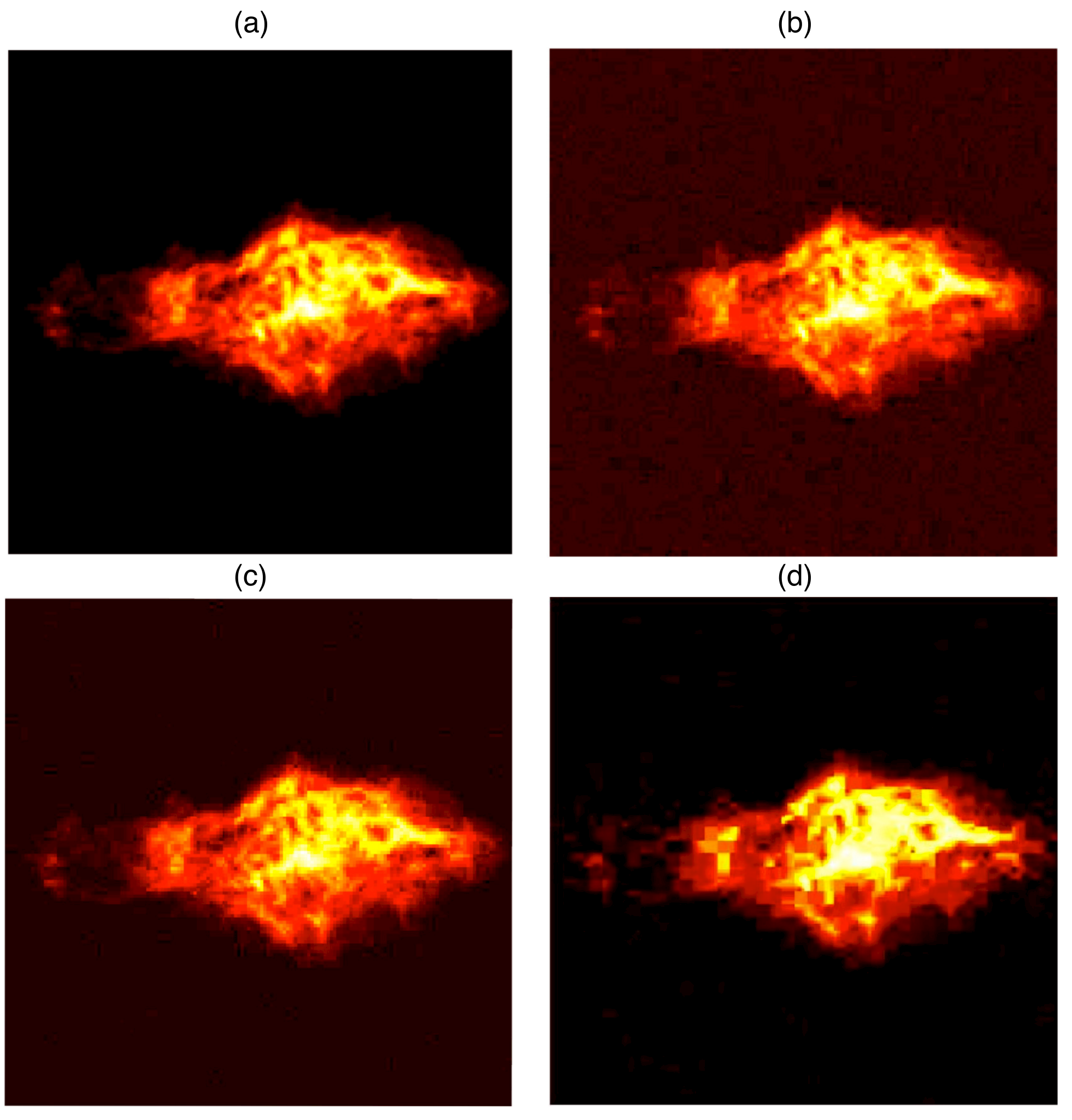}}
\caption{OMP reconstructions of the object (a), the radio emission from \object{3C58}, using the BDCT matrix as the sparsifying matrix. The object is of size 160 $\times$ 160 and measurements are collected from 100 antennas. Reconstructions are simulated with a URA in (b) (RE = 0.0082), an NRA of 0.14 in (c) (RE = 0.0028), and the VLA-based 'Y' array in (d) (RE = 0.0137). All pixel brightness values in the object and reconstructions are normalized by the object's maximum pixel brightness value, to lie on [0, 1]. Image (a) can be found at http://images.nrao.edu/529 (courtesy of NRAO/AUI and Michael Bietenholz, York University).}
\label{BDCT3C58}
\end{figure}

The OMP extended-source reconstructions in Figs. \ref{BDCTCrabNeb}, \ref{BDCTCassa}, \ref{BDCTM87}, and \ref{BDCT3C58} deviate from the MC results in intriguing ways. Most notably, in all cases the URA does not provide the most accurate reconstruction of the objects, despite providing the most incoherent measurement matrices. The URA's reconstructions in Figs. \ref{BDCTCrabNeb}b, \ref{BDCTCassa}b, \ref{BDCTM87}b, and \ref{BDCT3C58}b introduce high-frequency noise to the backgrounds of the galaxies, and also distort structures within the galaxies. The red- and yellow-colored internal structures of Fig. \ref{BDCTCrabNeb}a are broadened in Fig. \ref{BDCTCrabNeb}b; the central dark blue structure of Fig. \ref{BDCTCassa}a is highly distorted in Fig. \ref{BDCTCassa}b; and the large red structure on the right-hand side of Fig. \ref{BDCTM87}a is much diminished in Fig. \ref{BDCTM87}. In all cases, an NRA with an SD as low as 0.14 provides the most faithful reconstruction of the three arrays. Reinforcing the contradiction, the measurement matrix produced with this NRA had an MC of 0.28---more than twice the MC of the URA, 0.12. In the pixel basis the NRA could also emulate a URA, but only when the SD was great enough that the MC of the two arrays was also comparable. Reconstructions (not shown) of the sources \object{NGC2403}, \object{M33}, and \object{3C31} were also run, and demonstrated the same surprising trend: the NRA clearly provided the most faithful reconstructions with the smallest RE, despite giving much higher MC than the URA.

Similarly, despite giving the most coherent measurement matrices by far (see Fig. \ref{BDCTMC}), the 'Y' array provides reconstructions with smaller RE than the URA in Figs. \ref{BDCTCassa}d and \ref{BDCTM87}d. In Fig. \ref{BDCTCassa}, its reconstruction preserves the central dark blue structure of \object{Cassiopeia A} much more accurately than the URA, and in Fig. \ref{BDCTM87}d it preserves the large red structure on the right-hand side of \object{M87} that the URA's reconstruction fails to recover. The 'Y' array does suffer from an idiosyncratic distribution of error known as the blocking effect: when too few BDCT components are recovered by OMP, the missing components (typically high-frequency components) in each block create breaks in features that should span the blocks continuously. Despite the blocking effect, however, the key observation is that even when the 'Y' array does not outperform the URA (Figs. \ref{BDCTCrabNeb}d and \ref{BDCT3C58}d) its RE is surprisingly comparable given that its produces far more coherent measurement matrices (RE = 0.0752 vs. the URA's RE = 0.0633 in Fig. \ref{BDCTCrabNeb}, and RE = 0.0137 vs. the URA's RE = 0.0082 in Fig. \ref{BDCT3C58}). 

This discrepancy between the MC relationships and the extended source reconstructions has several possible roots---first, it is important to note that the calculation of MC is only mathematically synonymous to the calculation of the arrayÕs peak side-lobe in the natural pixel basis. With the BDCT (or any other transform) incorporated into the measurement matrix, the physical analogy to the peak side-lobe breaks down and the MC cannot be interpreted in the same way: though it is still theoretically relevant to array optimization through Eq. (\ref{Donoho2}), it no longer has any direct physical relevance. Many CS sampling studies have focused solely on MC or the RIP---just as many array studies have focused solely on side-lobes---but such indicators may have different implications depending on the sparsifying basis. Though the suitability of sparsifying bases for image reconstruction in radio interferometry was noted as early as in (Starck et al. 1994), we cannot fully appreciate CS in the field until we also understand how different sparsifying representations will affect array optimization.

We also note that, unlike with the pixel basis, the random perturbation of the YOPP array (not shown) does not significantly decrease the RE or blocking effect of the 'Y' array's reconstructions (unless the perturbation is so severe the original 'Y' framework is no longer recognizable). This is likely because the YOPP was designed to emulate a URA, the superior array for the pixel basis, rather than an NRA, the superior array for the BDCT: the largest baselines were perturbed, corresponding to the URA's coverage of larger baselines compared to the NRA. Developing a different modification scheme for the 'Y' array with the BDCT, such that it better emulates an NRA, is therefore of interest.

\section{Conclusions}
\label{Conc}
In the pixel basis, we showed that the URA optimizes the performance indicator MC and sets the benchmark in OMP reconstructions of both point and extended sources. Though this contradicts the general agreement on centrally condensed arrays in the well-sampled context, it is linked to previous results that arrays with small numbers of antennas should aim for a more uniform baseline distribution (Boone 2002; Holdaway 1997; Woody 2001a) so that the near-in side-lobes do not overpower the far side-lobes. The URA also showed promise in allowing for super-resolution in CS recovery, whereas the performance of all other tested arrays decayed rapidly in the under-resolved case.

When we used the BDCT to amplify object sparsity, despite resulting in significantly higher MC an NRA outperforms the URA to give the most faithful reconstructions of extended sources. This conveniently coincides with the preference for centrally condensed arrays in the well-sampled case. Such results also reveal the nearsightedness of solely analyzing MC, as is often done in optimizing sample distributions for CS, as this indicator does not preserve its physical meaning among different sparsifying bases. An array's MC is analogous to its peak side-lobe in the pixel and other spike-like bases, but cannot be interpreted the same way in other bases.

We also highlighted the YOPP array as a model for enhancing the VLA's 'Y' configuration with a deceptively trivial amount of randomization, as it emulates a URA in reconstructing both point and extended sources with the pixel basis. The principle of slight perturbation could apply to other patterned configurations as well, making them far more conducive to CS while largely preserving their practicality. This aligns with the early fundamental observation in (Lo 1964) that significantly less randomly spaced sensors than regularly spaced sensors are needed to achieve the same low level of side-lobes. In the context of CS, when we are constrained by a highly insufficient number of sensors, this principle becomes indispensable.

Our results have general implications and serve as an introductory look into arrays for CS in radio interferometry. Various factors can be studied further, in particular the sparsifying basis, which is critical to any application of CS.  That sparsifying bases can incorporate redundancy make them far more powerful than the pixel basis at representing natural images sparsely, and finding the optimal sparse representation of an object will clearly improve CS recovery performance (by the OMP error bounds in Eqs. (\ref{Donoho1}) and (\ref{Donoho2})). The NRA may be appropriate for the BDCT, but the BDCT is only one among many possible methods for sparsely representing images typical of radio astronomy. Array design with respect to wavelet transforms and frames, such as the isotropic undecimated wavelet transform (Starck et al. 2006), is of interest: unlike the BDCT, such multiresolution approaches avoid the issue of optimizing a block size, which depends on the scale of the structures being imaged. The CS recovery algorithm is also a key variable. OMP was chosen here because of its parallels to CLEAN, as well as its superior computational efficiency compared to the optimization-based BP and LASSO, which is critical in dealing with large data. However, these optimization approaches, as well as adaptations of the basic OMP framework, may respond to arrays differently. Also important is incorporating Earth-rotation aperture synthesis, as our results only target snapshot-mode interferometry, as well as running reconstructions on real rather than simulated data. As we look to deal with new floods of data from massive arrays, including the ALMA and the SKA, addressing these factors in array design for CS will become increasingly critical.

\begin{acknowledgements}
The author is grateful for financial support of the work from the Intel Science Talent Search, the Intel International Science and Engineering Fair, and the Junior Science \& Humanities Symposium.
\end{acknowledgements}

\end{document}